\documentclass[twocolumn,preprint2]{aastex63}

\newcommand{\kms}{km s$^{-1}$}
\newcommand{\kmsMpc}{{km s$^{-1}$ Mpc$^{-1}$}}

\newcommand{\profemail}{mglee@astro.snu.ac.kr}
\newcommand{\myemail}{yoojkim@astro.snu.ac.kr}
\accepted{September 30, 2020}
\submitjournal{ApJ}

\shorttitle{$H_0$ from Virgo Infall using TRGB Distances}
\shortauthors{Kim et al.}

\usepackage{color}
\usepackage{multirow}
\usepackage{amsmath}
\usepackage{textcomp}
\usepackage{comment}

\begin{document}

\title{
Determination of the Local Hubble Constant from Virgo Infall Using TRGB Distances
}

\author{Yoo Jung Kim, Jisu Kang, Myung Gyoon Lee}
\correspondingauthor{Myung Gyoon Lee} 
\affiliation{Astronomy Program, Department of Physics and Astronomy, Seoul National University, 1 Gwanak-ro, Gwanak-gu, Seoul 08826, Republic of Korea}
\email{\myemail,\profemail}
\author{In Sung Jang} 
\affiliation{ Leibniz-Institut  f{\"u}r Astrophysik Potsdam (AIP), An der Sternwarte 16, 14482, Potsdam, Germany}

\begin{abstract}
{
An independent determination of $H_0$ is crucial given the growing tension between the Hubble constant, $H_0$, derived locally and that determined from the modeling of the cosmic microwave background (CMB) originating in the early universe.
In this work, we present a new determination of $H_0$ using velocities and tip of the red giant branch (TRGB) distances to 33 galaxies located between the Local Group and the Virgo cluster.
We use a model of the infall pattern of the local Hubble flow modified by the Virgo mass, which is given as a function of the cosmological constants ($H_0$, $\Omega_\Lambda$), the radius of the zero-velocity surface $R_0$, and the intrinsic velocity dispersion, $\sigma_v$.
Fitting velocities and TRGB distances of 33 galaxies to the model, we obtain $H_0 =$ 65.8 $\pm$ 3.5($stat$) $\pm$ 2.4($sys$) \kmsMpc~ and $R_0 =$ 6.76 $\pm$ 0.35 Mpc. 
Our local $H_0$ is consistent with the global $H_0$ determined from CMB radiation, showing no tension.
In addition, we present new TRGB distances to NGC 4437 and NGC 4592 which are located near the zero-velocity surface: D = 9.28 $\pm$ 0.39 Mpc and D = 9.07 $\pm$ 0.27 Mpc, respectively.
Their spatial separation is 
0.29$^{+0.30}_{-0.03}$ Mpc, 
suggesting that they form a physical pair.

} 
\end{abstract}
\keywords{Distance indicators (394), Stellar distance (1595), Galaxy stellar halos (598), Galaxy stellar content (621), Virgo Cluster (1772), Cosmology (343), Hubble constant (758), Cosmological parameters (339)}


\section{Introduction}\label{sec_introduction}

The Hubble tension is the discrepancy in recent cosmology that the Hubble constant ($H_0$) determined locally is significantly larger than the global value determined by observation of cosmic microwave background radiation (CMB) by \citet{Planck18}, $H_0 =$ 67.4 $\pm$ 0.5 \kmsMpc~ (\citet{rie19b,ver19}, and references therein).
The tension has been emphasized since \citet{rie16} obtained $H_0 =$ 73.24 $\pm$ 1.74 \kmsMpc~ based on an empirical Cepheid calibration of the absolute magnitudes of Type Ia supernovae in the more distant Hubble flow.
As refinements to the techniques have been made, the measurement uncertainties have shrunk but the range of measured values have not, yielding $H_0 =$ 74.03 $\pm$ 1.42 \kmsMpc~ recently \citep{Riess19}.
This is 4.4$\sigma$ discrepant with the Planck result and can be interpreted as a significant disagreement between the measurements of $H_0$ from the early universe and from the late universe.

In order to understand the cause of this discrepancy, we must take a closer look at these different techniques.
In particular, because the value of $H_0$ based on standard candles depends highly on the zero-point of the distance scale, another independent distance ladder aside from Cepheids is crucial.
Thus, there has been much effort to measure $H_0$ independently using the tip of the red giant branch (TRGB) \citep{lee93} method.

The TRGB is the truncation of the red giant branch (RGB) sequence that corresponds to the helium flash point in the evolutionary stages of low-mass stars.
Thus, the TRGB appears as an edge of the RGB sequence in color magnitude diagrams (CMDs) and the luminosity of the TRGB varies only slightly with metallicity in $I$ band.
Therefore, the TRGB provides a robust standard candle \citep{lee93,jan17,jan17b,lee18,freed19,freed20}.

\citet{jan17b}(TIPSNU) obtained $H_0 =$ 71.17 $\pm$ 1.66($stat$) $\pm$ 1.87($sys$) \kmsMpc~ by calibrating the luminosity of SNe Ia using TRGB distances to galaxies hosting SNe Ia.
Their $H_0$ value weakened the Hubble tension as their value lies between those measured by the Cepheid-calibrated SNe Ia method and those measured from CMB analysis. 
Later, \citet{freed19} (Carnegie-Chicago Hubble Program; CCHP) reduced the uncertainty using a larger sample, yielding $H_0 =$ 69.8 $\pm$ 0.8($stat$) $\pm$ 1.7($sys$) \kmsMpc~ (revised to 69.6 $\pm$ 0.8($stat$) $\pm$ 1.7($sys$) in \citet{freed20}). This value of the Hubble constant agrees at the 1.2$\sigma$ level with that from the Planck Collaboration and at the 1.7$\sigma$ level with that from the Cepheid distance scale.

To date, most measurements of the local $H_0$ use distant galaxies hosting supernovae in order to minimize the effects of peculiar velocities. 
However, galaxies located between the Local Group and the Virgo Cluster ($D=$ 16.5 Mpc \citep{kas20}) can also be used to obtain $H_0$, if we simultaneously consider their peculiar motion and the Hubble flow.
The local Hubble flow is modified by the gravity of the Virgo cluster, showing our infall pattern toward the center of the Virgo cluster.
\citet{lyn81} and \citet{san86} suggested that accurate distances and velocities to outlying members of the Local Group can be used to determine the age of the universe and the mass of the Local Group, based on the timing argument \citep{kah59}. 
Their model was based on the Lema\^{i}tre-Tolman-Bondi model, which describes the dynamics of a pressure-free spherically symmetric system of particles \citep{lem33, tol34, bon47}.
\citet{pei06, pei08} modified the model by including the cosmological constant, $\Omega_\Lambda$, and applied it to the Virgo infalling galaxies.
Fitting 27 galaxies with Tully-Fisher distances to the theoretical velocity-distance relation, they obtained $H_0 =$ 71 $\pm$ 9 \kmsMpc.

In this work, we use TRGB distances to galaxies between the Local Group and the Virgo cluster to determine $H_0$ with reduced uncertainty.
The large uncertainty of $H_0$ in \citet{pei08} is due to the large uncertainty of Tully-Fisher distance measurements, typically 15\%.
Given that uncertainties of TRGB distances are typically 5\%, the use of TRGB distances to fit the velocity-distance relation is expected to yield a more precise determination of $H_0$.

Fortunately, TRGB distances to many galaxies in front of the Virgo cluster are available today.
In particular, \citet{kar14, kar18} observed 29 galaxies with Hubble Space Telescope (HST) and derived TRGB distances to these galaxies in order to investigate Virgo infall and estimate the dynamical mass within zero-velocity radius.
Moreover, \citet{tik20} obtained TRGB distances to 18 
additional 
galaxies from archival {\it HST} images and noted that six of them are projected onto the Virgo cluster.

Adding to these samples, we determine TRGB distances to two galaxies in the southern Virgo infall region using archival \textit{HST} data: NGC~4437 and NGC~4592. Hyper Suprime-Cam (HSC) color images of these galaxies \citep{aih19} are shown  in Figure \ref{fig_FOV}.
NGC~4437 (= NGC~4517) is an edge-on spiral galaxy and 
the brightest 
galaxy located beyond the Local Sheet in \citet{kar14, kar18} samples.
Its TRGB distance is measured to be $8.34 \pm 0.83$~Mpc by \citet{kar14}.
NGC~4592 is a dwarf spiral galaxy located 1.8\textdegree~ away from NGC~4437 
in the sky, 
but its TRGB distance has not been studied.
The Tully-Fisher distance 
to NGC 4592 is $11.60 \pm 2.30$~Mpc \citep{sor14} and thus has not been included in the Local Volume (D $<$ 11 Mpc) galaxies sample defined by \citet{kar13}.

This paper is structured as follows. 
In Section \ref{sec_data}, we describe the \textit{HST} data and the reduction process used for the TRGB measurement. 
In  \S\ref{sec_TRGB}, we describe the methods and results of TRGB distance measurements. 
In \S\ref{sec_virgo_infall}, we illustrate on the Virgo infall fitting and determination of $H_0$.  
In \S\ref{sec_group}, we discuss the physical separation between NGC~4437 and NGC~4592. 
In \S\ref{sec_err}, we examine possible systematic uncertainties in our $H_0$ measurement.
In \S\ref{sec_H0}, we compare our determination of $H_0$ with the ones from other studies. 
In \S\ref{sec_summary}, we summarize our main results.


\begin{deluxetable*}{lccccccc}[hbt!]
\tabletypesize{\footnotesize}
\tablewidth{0pt} 

\caption{A Summary of {\it HST} Observations for NGC~4437 and NGC~4592 Fields }\label{tab-HST}

\tablehead{
\colhead{Field} 	& \colhead{R.A.} & \colhead{Decl.} & \colhead{Instrument} & \multicolumn3c{Exposure Time} &  \colhead{Prop. ID} \\
   &  \colhead{(J2000)} & \colhead{(J2000)}   &   &   \colhead{F555W ($V$)} & \colhead{F606W ($V$)} & \colhead{F814W ($I$)} & }
\startdata
NGC~4437 	& 12:32:46.83	& +00:05:01.1	& ACS/WFC &  & 1640s & 2076s & 12878 \\
NGC~4592 & 12:39:10.54	& --00:31:46.5 & WFC3/UVIS & 5530s & & 5674s & 11360 \\
\enddata

\end{deluxetable*}


\section{Data and Data Reduction}\label{sec_data}

\begin{figure}[hbt!]
\centering
\includegraphics[scale=1.0]{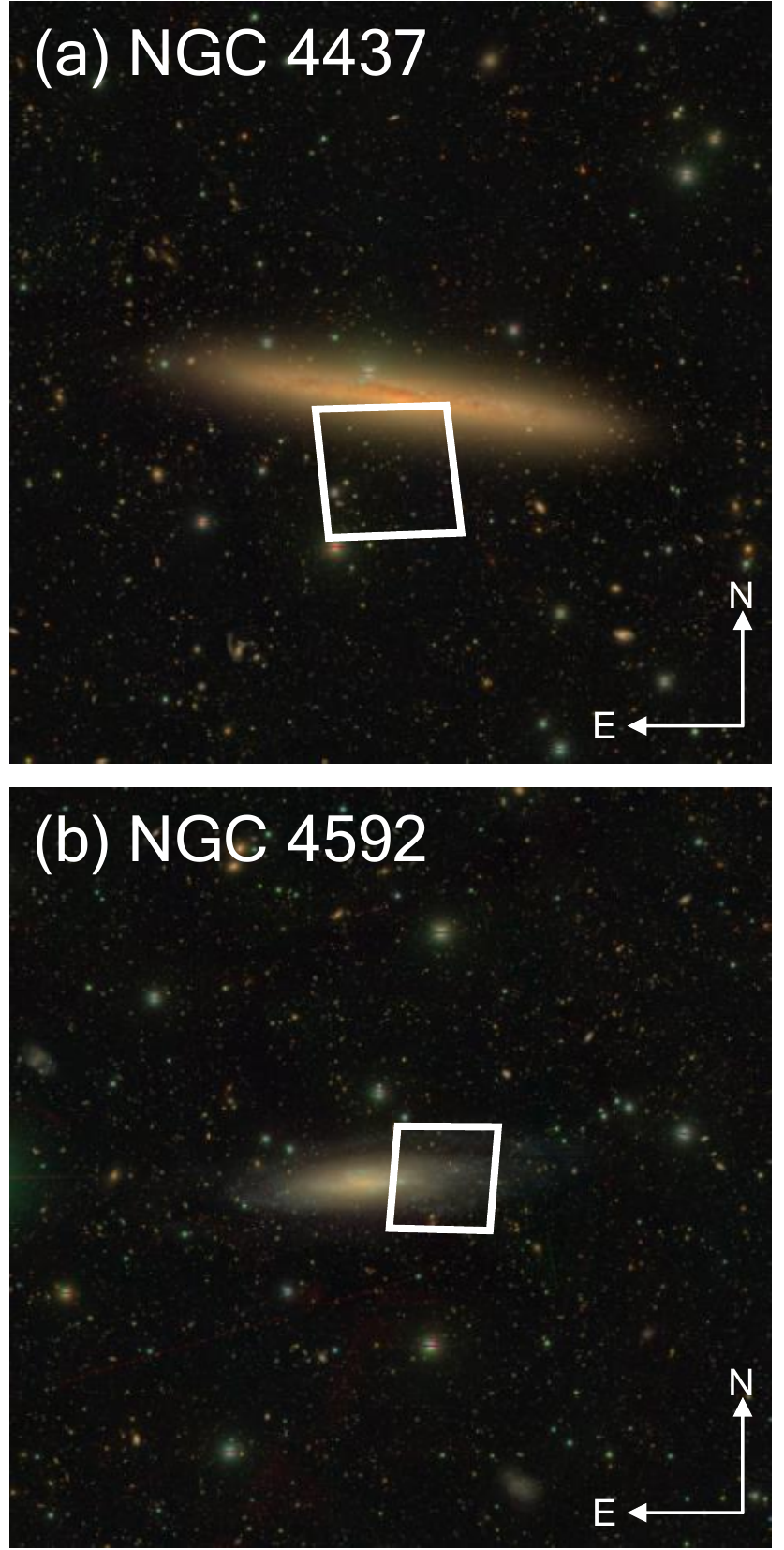} 
\caption{Hyper Suprime-Cam(HSC) {\it g, r, i} color images of (a) NGC~4437 (b) and NGC~4592 \citep{aih19}. The field of view is 20$\arcmin$ $\times$ 20$\arcmin$. 
$HST$ fields taken with ACS (a) and WFC3 (b) are shown by white rectangles. 
}
\label{fig_FOV}
\end{figure}

We use two fields of archival {\it HST} images to measure TRGB distances to NGC~4437 and NGC~4592, as shown by rectangles in Figure \ref{fig_FOV}. The {\it HST} observations are summarized in Table \ref{tab-HST}. 
The NGC~4437 minor axis halo field was taken with {\it HST/ACS} F606W and F814W filters, and the NGC~4592 major axis halo field was observed with {\it HST/WFC3} F555W and F814W filters.
Individual \texttt{.flc} (charge transfer efficiency corrected) frames of each filter 
were aligned using \texttt{TWEAKREG} then were combined using \texttt{ASTRODRIZZLE} \citep{gon12}.

For the detection and photometry of the point sources, we used the most recent version of \texttt{DOLPHOT} \citep{dol00}. 
We 
used %
the \texttt{Fitsky} = 2 option for the local sky estimation, which is applicable for crowded field photometry.
We applied the selection criteria of point sources for both bands as follows: 
--0.3 $<$ \texttt{SHARP} $<$ 0.3, 
\texttt{CROWD} $<$ 0.25, 
\texttt{CHI} $<$ 2.5, 
\texttt{ROUND} $<$ 1,
\texttt{flag} $<$ 4, and
\texttt{S/N} $>$ 3. 
We used the Vega magnitude system following the \texttt{DOLPHOT} output parameters.


\begin{figure}[hbt!]
\centering
\includegraphics[scale=0.57]{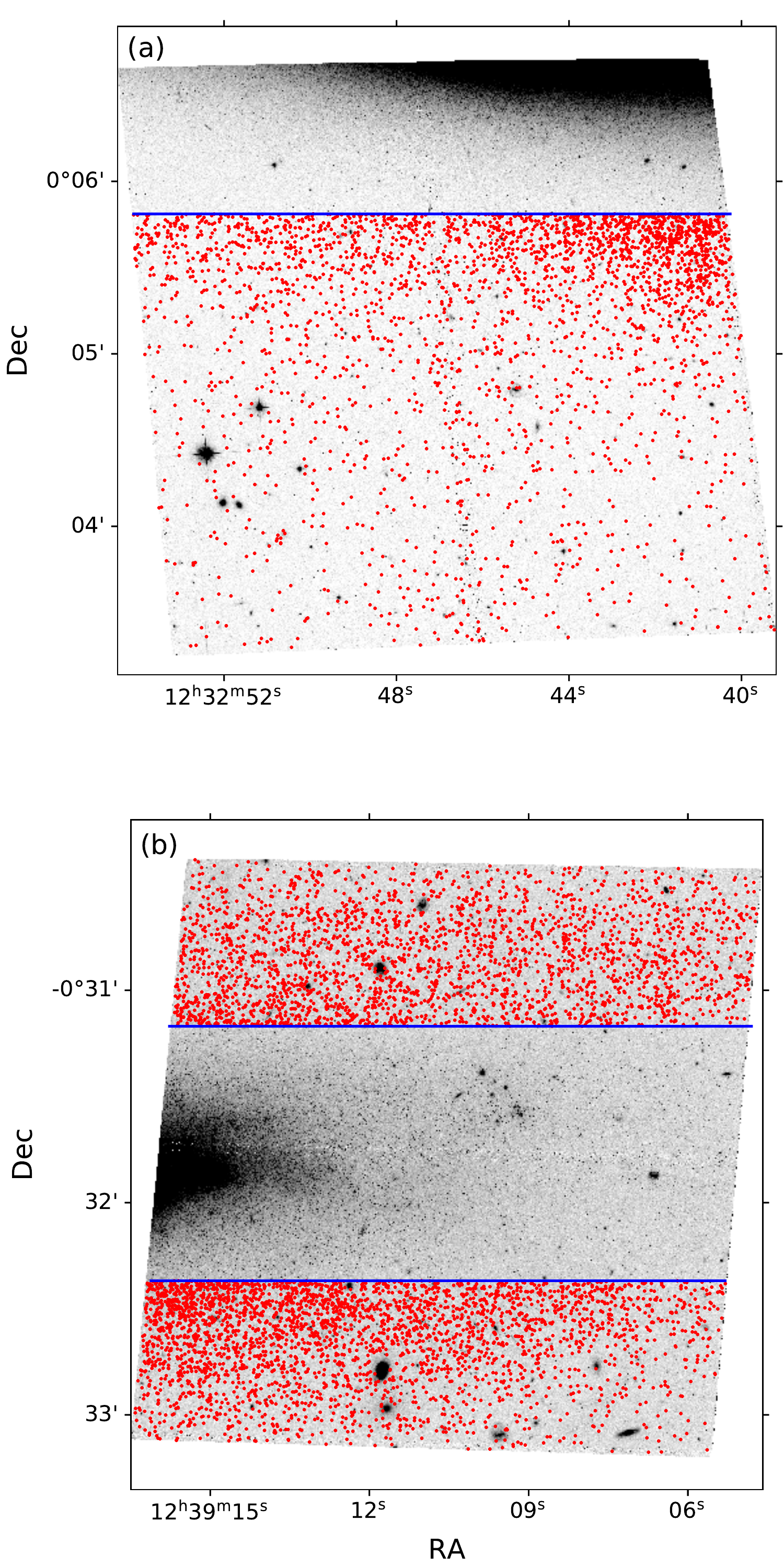} 
\caption{F814W {\it HST} images of (a) NGC~4437 and (b) NGC~4592. Blue solid lines denote boundaries of the spatial selection criteria. 
Red dots represent selected RGB star candidates from the shaded regions in Figure~\ref{fig_TRGB}. 
}
\label{fig_star_selection}
\end{figure}


\begin{figure}[hbt!]
\centering
\includegraphics[scale=0.55]{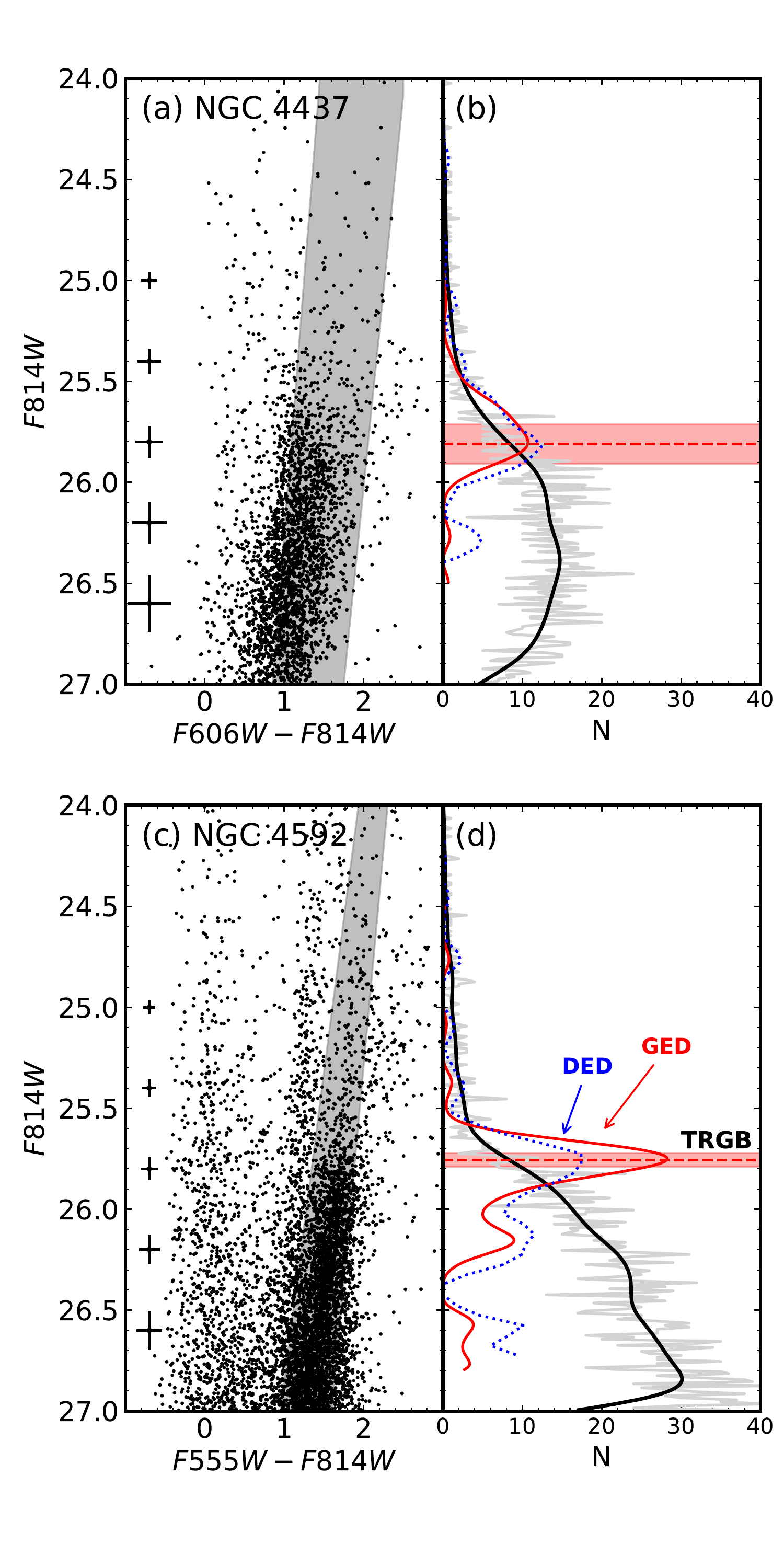} 
\caption{(Left) CMDs of spatially selected resolved stars in (a) NGC 4437 and (c) NGC 4592. Stars in the gray shaded region are selected as RGB star candidates. Mean photometric magnitude errors of RGB stars as a function of magnitude are shown as crosses on the left. (Right) LFs of selected RGB stars are shown as light gray lines and GLOESS-smoothed LFs are represented as black solid lines. 
Edge responses to the Sobel filter [--1, 0, +1] applied to GLOESS-smoothed LFs are shown as red solid lines (GLOESS edge detection; GED) and corresponding TRGBs are marked as red dashed lines. Edge responses of the direct edge detection (DED) method are plotted as blue dotted lines. 
}
\label{fig_TRGB}
\end{figure}


\begin{figure}[hbt!]
\centering
\includegraphics[scale=0.57]{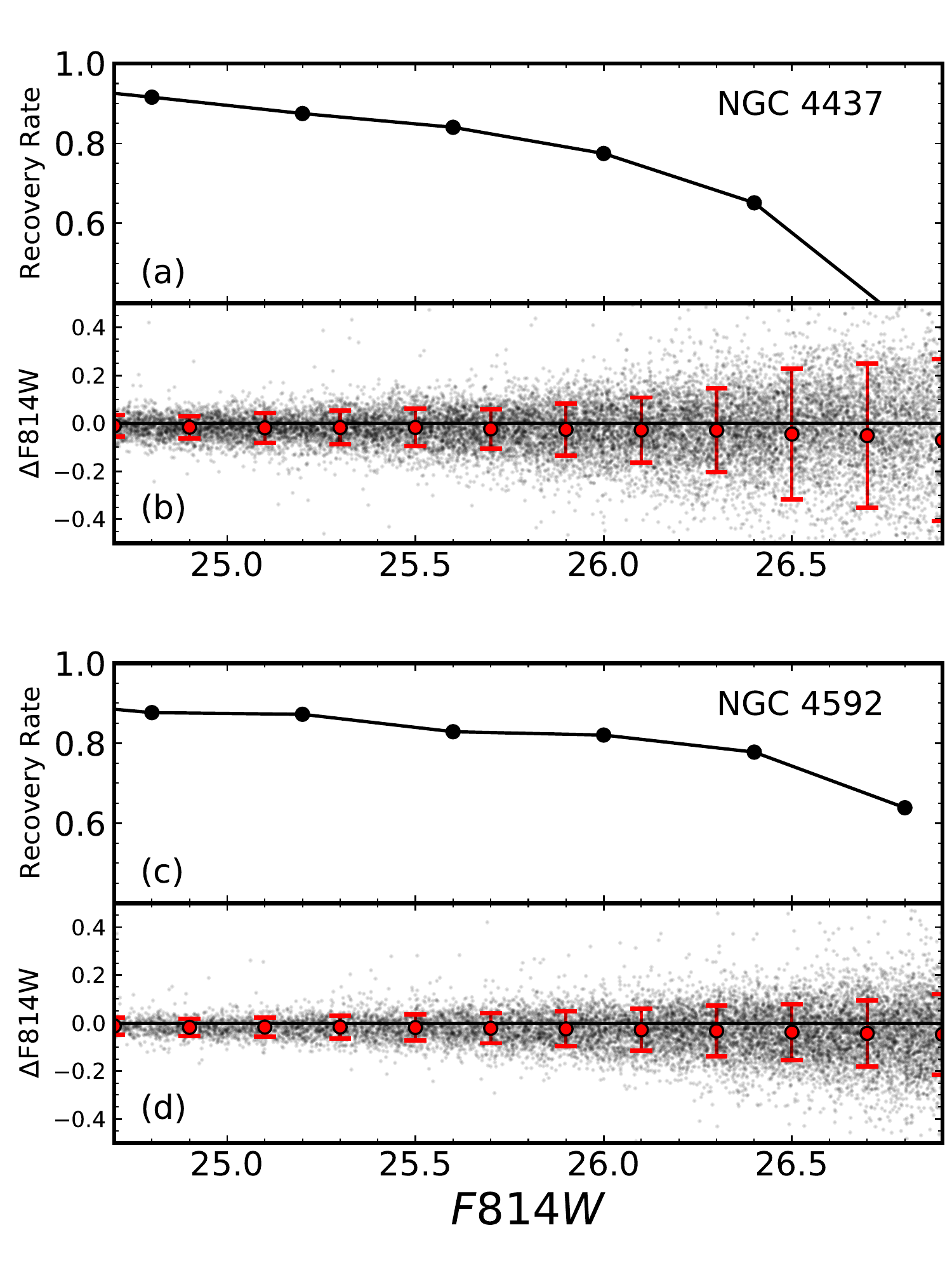} 
\caption{Panels (a) and (c): RGB star completeness derived from artificial star tests for NGC 4437 and NGC 4592, respectively. Panels (b) and (d): (Input mag -- output mag) of artificial stars (gray dots). 
Red circles and error bars represent mean values and mean errors, respectively.} 
\label{fig_artstar}
\end{figure}

\section{TRGB Distances to NGC~4437 and NGC~4592}\label{sec_TRGB}

In this Section, we first describe our selection criteria for RGB star candidates.
Then we show two different approaches to measure the tip: detection of edge response using Sobel filter \citep{lee93} and maximum likelihood method \citep{men02, maka06}.
Next, we summarize our results and compare the TRGB detection methods.
Lastly, we compare our TRGB distance to NGC~4437 with previous measurements.

\subsection{Selection of RGB Star Candidates}

We use spatial and color constraints 
to 
select RGB star candidates in the galaxies. 
We use stars in the outer region of the {\it HST} fields 
to avoid contamination from 
younger stellar populations 
and 
high stellar 
crowding as much as possible, as marked by blue solid lines in Figure \ref{fig_star_selection}.
Since NGC~4437 is an edge-on disk galaxy, 
and the ACS field is shifted off the disk plane, 
halo stars 
can be 
effectively selected using spatial constraints.
On the other hand, NGC~4592 was observed with WFC3, which has a smaller field of view, and the field lies at the center of the disk plane, so the outer regions of the NGC~4592 field may include a larger fraction of disk stars compared with NGC~4437. 

CMDs of the selected stars are displayed in the left panels of Figure \ref{fig_TRGB}.
The RGB sequence is seen clearly in the color range $1.1 < F606W - F814W < 1.7$ for NGC 4437 and $1.5 < F555W - F814W < 1.9$ for NGC 4592.
For NGC 4592, vertical sequences of blue and red young stars are conspicuous around the color $F555W - F814W \sim 0$ and $F555W - F814W \sim 1.2$ (blue ones are blue supergiants and main-sequence stars,
and red ones are red supergiants).
Therefore, in order to minimize the contribution from young stars and other non-RGB sources, we selected RGB star candidates from the shaded regions.
The selected stars are expected to be dominated by 
old 
RGB stars and a few asymptotic giant branch (AGB) stars.
The spatial locations of these stars are shown as red dots in Figure \ref{fig_star_selection}.

Finally, we conduct artificial star tests for RGB stars using \texttt{DOLPHOT}.
We generate 50,000 artificial stars that satisfy the 
color and spatial constraints described above, in the magnitude range 
from $F814W =$ 24 to 27~mag. 
Figure \ref{fig_artstar} shows the results of artificial star tests.
The completeness (a number ratio of recovered stars to added stars) as a function of input magnitude are shown in panels (a) and (c).
The completeness reaches 50\% level at about $F814W =$ 26.6 mag for NGC~4437 and about $F814W =$ 27.0 mag for NGC~4592.
The 
input magnitude minus output magnitude 
of each star is shown in panels (b) and (d). The mean errors are smaller than 0.03 mag in the range of $F814W <$ 26 mag.

\subsection{Edge Detection Method}

The basic concept of 
the 
edge detection method is an identification of the peak of the edge response, which is calculated by convolving  the luminosity function (LF) of RGB stars with a Sobel kernel \citep{lee93}. 
There have been a number of refinements and variations since then (see \citet{hatt17} and \citet{jan18} for a comparison of recent methods). 
We use two of the recent methods in this study: (1) edge detection for smoothed LFs \citep{hatt17} and (2) the direct edge detection method \citep{jan17b}.

In \citet{hatt17}, a finely binned (0.01 mag) LF of selected RGB candidates is smoothed using a Gaussian-windowed, locally weighted scatterplot smoothing (GLOESS), with the characteristic width of Gaussian weighting, $\sigma_s$. 
Then the Sobel kernel [--1, 0, +1] is applied to the smoothed LF, and the magnitude of maximum edge response corresponds to the TRGB. 
This smoothing technique is introduced in order to suppress false edges \citep{hatt17}.
To find the optimal smoothing scale ($\sigma_s$) that minimizes the sum of systematic and random error, TRGB detection by the artificial star luminosity function (ASLF) was conducted (see \citet{hatt17} for details).
We followed the same method and found optimal $\sigma_s$ to be 0.10 mag and 0.08 mag for NGC~4437 and NGC~4592, 
respectively. 

Figure \ref{fig_TRGB} shows finely binned LFs (light gray lines), smoothed LFs (black solid lines), and edge responses (red solid lines) for these two galaxies.
The maximum edge response is seen at $F814W_{{\rm TRGB}} \approx 25.8$ mag for both galaxies. 
For NGC~4437, the systematic error (input TRGB -- output TRGB) obtained from the ASLF test is $\Delta\mu =$ 0.076 mag and the random error is $\sigma =$ 0.055 mag.
Similarly, for NGC~4592, $\Delta\mu =$ 0.011 mag and $\sigma =$ 0.020 mag.
Moreover, artificial star tests show that there are slight systematic magnitude offsets (input mag -- output mag) at the TRGB magnitude: $\Delta F814W = -0.025$ mag for NGC~4437 and $\Delta F814W = -0.024$ mag for NGC~4592.
These magnitude offsets are combined into systematic errors.
Considering the uncertainties mentioned above, the TRGB magnitude of NGC~4437 is derived to be 
$F814W_{{\rm TRGB}} =$ 25.811 $\pm$ 0.080 ($sys$) $\pm$ 0.055 ($stat$) mag and that of NGC 4592 is $F814W_{{\rm TRGB}} =$ 25.756 $\pm$ 0.026 ($sys$) $\pm$ 0.020 ($stat$) mag.
These results are summarized in Table~\ref{tab-TRGB}.

Next, we try the direct edge detection method described in \citet{jan17b}. 
We apply a Sobel filter $[-1, -2, -1, 0, +1, +2, +1]$ 
directly to the LFs derived with a 0.05 mag bin.
The edge response thus derived is shown as dotted blue lines in Figure~\ref{fig_TRGB}. The maximum edge responses are seen at magnitudes similar to those from  the GLOESS method.
The uncertainty of the TRGB magnitude is obtained by bootstrap resampling RGB stars 10,000 times.
Combining these errors with the systematic magnitude offsets given above, we derive $F814W_{{\rm TRGB}} =$ 5.829 $\pm$ 0.049 mag for NGC~4437 and $F814W_{{\rm TRGB}} =$ 25.763 $\pm$ 0.050 mag for NGC~4592, which are in excellent agreement with those of the GLOESS method. 

We use the recent TRGB calibration of \citet{jan20} to obtain distances.
\citet{jan20} used the megamaser-based distance to NGC~4258 \citep{rei19} as an anchor and obtained $M_{F814W}^{TRGB} = -4.050 \pm 0.028 (stat) \pm 0.048 (sys)$ in the ACS magnitude.
This is applicable to the blue TRGB ($(F606W - F814W)_{0} < 1.5$ or $(F555W - F814W)_{0} < 2.1$), where $I$-band TRGB remains approximately constant with color.
Mean TRGB colors of both 
NGC~4437 ($(F606W - F814W)_{0, {\rm TRGB}} =$ 1.20 $\pm$ 0.10) and 
NGC~4592 ($(F555W - F814W)_{0, {\rm TRGB}} =$ 1.65 $\pm$ 0.06) 
satisfy these criteria.
Since the calibration is obtained in $F814W_{{\rm ACS}}$ magnitude system, we convert $F814W_{{\rm WFC3}}$ magnitude to $F814W_{{\rm ACS}}$ magnitude for NGC 4592 using the empirical transformation derived by \citet{jan15}.
Here, we correct the apparent TRGB magnitudes for the foreground extinction: 
$A_{{\rm F606W (ACS)}}=0.059$ and $A_{{\rm F814W (ACS)}}=0.036$ for NGC~4437 and 
$A_{{\rm F555W (WFC3)}}=0.055$ and $A_{{\rm F814W (WFC3)}}=0.034$ for NGC~4592 \citep{schlegel98, sch11}.
Then the resulting distances for the GLOESS method are 9.22 $\pm$ 0.23 ($stat$) $\pm$ 0.41 ($sys$) Mpc for NGC~4437 and 9.03 $\pm$ 0.08 ($stat$) $\pm$ 0.25 ($sys$) Mpc for NGC 4592.
For direct edge detection method, the resulting distances are 9.30 $\pm$ 0.18 ($stat$) $\pm$ 0.26 ($sys$) Mpc for NGC 4437 and 9.06 $\pm$ 0.18 ($stat$) $\pm$ 0.25 ($sys$) Mpc for NGC 4592.
These results are summarized in Table \ref{tab-TRGB}.
In Table \ref{tab-TRGB}, we also show the distances calibrated by \citet{freed20}, which used the geometric distance to the Large Magellanic Cloud as an anchor to TRGB 
($M_{F814W}^{TRGB} = -4.054 \pm 0.022 (stat) \pm 0.039 (sys)$). 
It is noted that two calibrations agree well within 1\%.

\subsection{Maximum Likelihood TRGB Detection Method}

We use the maximum likelihood TRGB detection method introduced by \citet{men02} and \citet{maka06}. This method fits an observed LF to a model with four free parameters (TRGB magnitude, RGB slope $a$, AGB slope $c$, and discontinuity $b$):
\begin{equation}
\psi=
\begin{cases}
10^{a(m-m_{TRGB})+b},\quad m - m_{TRGB} \geq 0\\
10^{c(m-m_{TRGB})},\quad \quad m - m_{TRGB} < 0. 
\end{cases}
\end{equation}
We obtain a smoothed model by convolving the above model with photometric uncertainty and completeness from artificial star tests and use it for fitting.
Sobel edge detection results derived above are used as initial guesses.

\begin{figure}[hbt!]
\centering
\includegraphics[scale=0.57]{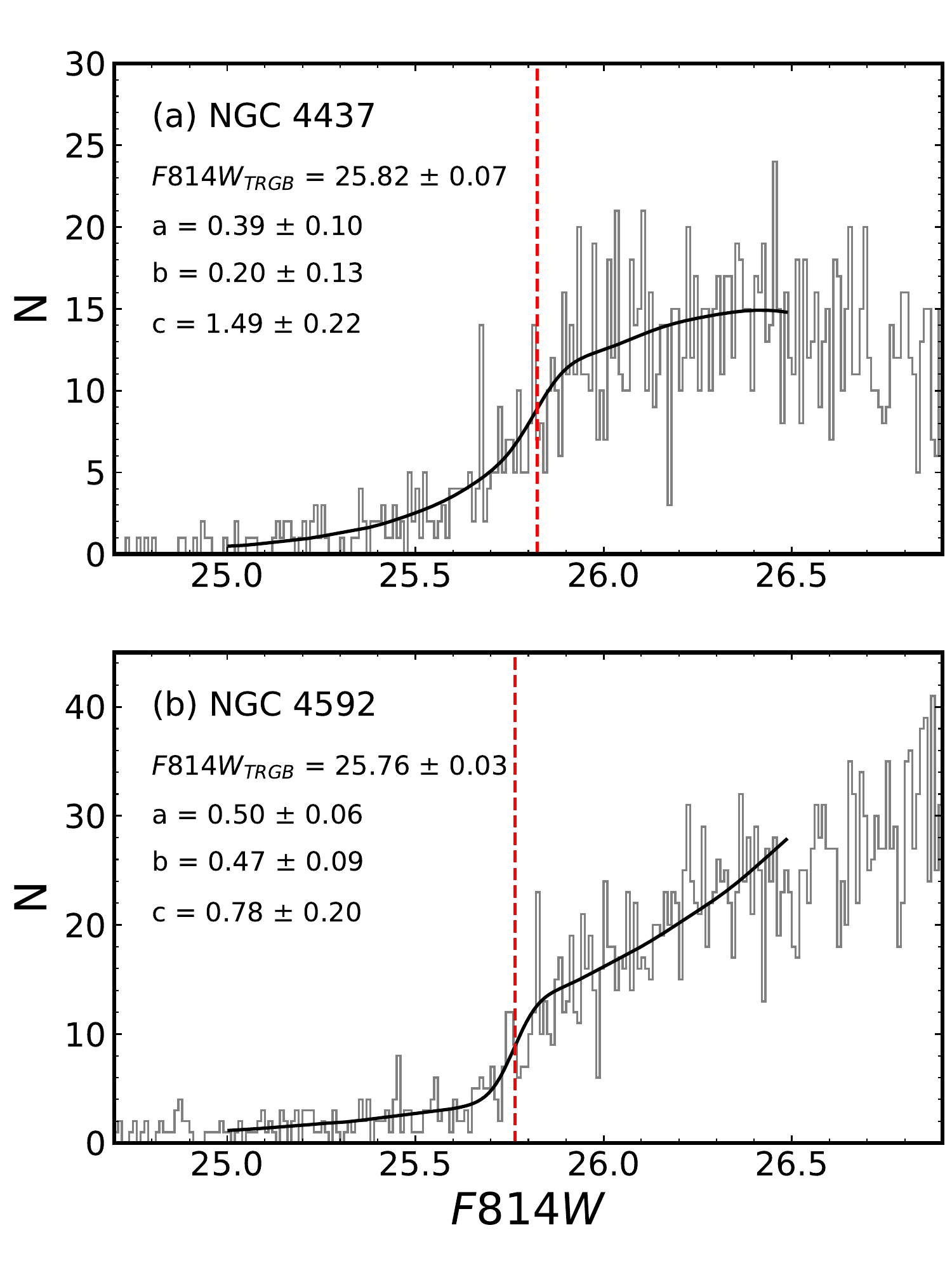} 
\caption{Results of maximum likelihood TRGB detection. Gray bars show 0.01 mag binned LF and black solid lines represent model LF convolved with photometric errors and incompleteness, 
$\varphi$($m_{{\rm TRGB}}$, a, b, c). 
Red vertical lines mark the TRGB. 
}
\label{fig_ML}
\end{figure}

Figure~\ref{fig_ML} shows the results of the maximum likelihood optimization.
Standard errors of the parameters are calculated by the square root of the inverse Hessian matrix.
For NGC~4592, the slopes on both sides 
($a=0.50\pm0.06$, $c=0.78\pm0.20$) are well-defined and the discontinuity is clear ($b=0.47\pm0.09$).
In contrast, for NGC~4437, 
the AGB slope ($c=1.49\pm0.22$) is much larger and the discontinuity ($b=0.20\pm0.13$) is smaller compared to those of NGC~4592, showing a more ambiguous tip.  
Combining the fitting errors with the systematic magnitude offsets given above, we derive the TRGB magnitude 
to be $F814W_{{\rm TRGB}} = 25.824 \pm 0.072$ mag  for NGC~4437 
and $F814W_{{\rm TRGB}} =25.765 \pm 0.034$ mag for NGC~4592.
Applying \citet{jan20} TRGB calibration, we derive distances: 
9.28 $\pm$ 0.29 ($stat$) $\pm$ 0.26 ($sys$) Mpc for NGC~4437 and 
9.07 $\pm$ 0.10 ($stat$) $\pm$ 0.25 ($sys$) Mpc for NGC~4592 (Table~\ref{tab-TRGB}).
These distance estimates are almost the same as those based on the \citet{freed20} calibration, as listed in Table~\ref{tab-TRGB}.


\subsection{
 Summary of TRGB Distance Estimation} 

We used three different methods to obtain the TRGB distances of NGC~4437 and NGC~4592.
Table \ref{tab-TRGB} lists a summary of TRGB distance measurements derived from the three methods.
Here we briefly compare the three methods.
Since the observed LFs of NGC~4437 and NGC~4592 are different, we compare each method for each of the two cases.

The LF of NGC~4437 does not show a sharp edge due to the large slope of the AGB LF and a weak TRGB discontinuity. This makes the parameterization of the LF difficult.  For instance, there were several local maxima when using the maximum likelihood method. Thus we tried the fitting with a number of initial conditions to find the true maximum. For the same reason, it was difficult to construct an ASLF that resembles the observed LF when using the GLOESS edge detection method. Since the TRGB discontinuity derived in the maximum likelihood method is small, we set a large ratio of AGB stars to RGB stars in the ASLF. 
This might have resulted in a larger uncertainty value for the GLOESS edge detection method. 
Compared to the complexity of these methods, the direct edge detection method is the simplest.
However, a systematic uncertainty might be introduced because of a fixed bin size of the LF.
While the GLOESS edge detection method estimates the systematic and random uncertainties using the ASLF and finds an optimal smoothing scale that minimizes the quadratic sum of systematic and random errors, the direct edge detection method does not include this procedure.

On the other hand, the LF of NGC~4592 clearly shows a sharp edge. The maximum likelihood method stably finds the maximum, and the ASLF is easily constructed. However, as seen in Figure \ref{fig_TRGB}(d), a false peak exists at $F814W \approx 26.1$ mag. While the GLOESS method suppresses this false edge, the direct edge detection method does not. Thus, the uncertainty of the direct edge detection method is slightly larger than those of the other two methods.

In conclusion, the TRGB distance measurements derived from the three methods coincide well within the error range. It does not matter which of the three is used, but we adopt the results of the maximum likelihood method for the following analysis. This is because TRGBs of most galaxies in our Virgo infall sample (\S\ref{sec_virgo_infall}) are obtained with the maximum likelihood method. TRGBs of only five galaxies in the \citet{tik20} sample are measured with the edge detection method used in \citet{lee93}. All the others are derived using the maximum likelihood method. Therefore, in order to be consistent, we chose the results of the maximum likelihood method.


In addition, our results show that distances to NGC~4437 and NGC~4592 are very similar.
The spatial separation between the two is discussed in Section~\ref{sec_group}.

\subsection{Comparison with Previous TRGB Magnitude for NGC 4437}

\citet{kar14} used the maximum likelihood analysis for TRGB detection and obtained $F814W_{{\rm TRGB}} =$ 25.65$^{+0.16}_{-0.12}$ mag for NGC 4437.
The resulting distance is 10\% smaller than our measurement but is consistent considering the large error range presented.
As discussed above, NGC 4437 shows an ambiguous tip in the maximum likelihood analysis.
However, the edge response function we derived in Figure \ref{fig_TRGB} shows a blunt but single peak at $F814W_{ {\rm TRGB}} \sim 25.8$ mag, showing no peak at the TRGB magnitude proposed by \citet{kar14}.
Therefore, we adopt our value for NGC 4437 for the following analysis.


\begin{deluxetable*}{lcccc}[htb!]
\tabletypesize{\footnotesize}
\tablewidth{0pt} 

\caption{A Summary of TRGB Distances}\label{tab-TRGB}
\tablehead{
\colhead{Galaxy}	& \multicolumn{2}{c}{NGC~4437} & \multicolumn{2}{c}{NGC~4592} }
\startdata
TRGB Color & \multicolumn{2}{c}{$(F606W - F814W)_{{\rm ACS}} = 1.23 \pm 0.10$} & \multicolumn{2}{c}{$(F555W - F814W)_{{\rm WFC3}} = 1.68 \pm 0.06$} \\ \hline 
TRGB Magnitudes & \multicolumn{2}{c}{$F814W_{{\rm ACS}}$} & \multicolumn{2}{c}{$F814W_{{\rm WFC3}}$} \\ 
\hline 
GLOESS Edge Detection \citep{hatt17}& \multicolumn{2}{c}{25.811 $\pm$ 0.097} & \multicolumn{2}{c}{25.756 $\pm$ 0.033} \\ 
Direct Edge Detection \citep{jan17b}& \multicolumn{2}{c}{25.829 $\pm$ 0.049} & \multicolumn{2}{c}{25.763 $\pm$ 0.050} \\ 
Maximum Likelihood \citep{maka06}& \multicolumn{2}{c}{25.824 $\pm$ 0.072} & \multicolumn{2}{c}{25.765 $\pm$ 0.034} \\ \hline 
& \multicolumn{4}{c}{Calibration} \\ \cline{2-5}
Distances & \citet{jan20} & \citet{freed20} & \citet{jan20} & \citet{freed20} \\ \hline 
GLOESS Edge Detection \citep{hatt17}& 9.22 $\pm$ 0.47 & 9.24 $\pm$ 0.46& 9.03 $\pm$ 0.27 & 9.04 $\pm$ 0.23  \\ 
Direct Edge Detection \citep{jan17b}& 9.30 $\pm$ 0.32 & 9.32 $\pm$ 0.29& 9.06 $\pm$ 0.31 & 9.07 $\pm$ 0.28  \\ 
Maximum Likelihood \citep{maka06}& 9.28 $\pm$ 0.39 & 9.30 $\pm$ 0.36 & 9.07 $\pm$ 0.27 & 9.08 $\pm$ 0.24  \\ 
\enddata 
\end{deluxetable*}



\section{Virgo Infall}\label{sec_virgo_infall}

\begin{figure} 
\centering
\includegraphics[scale=0.45]{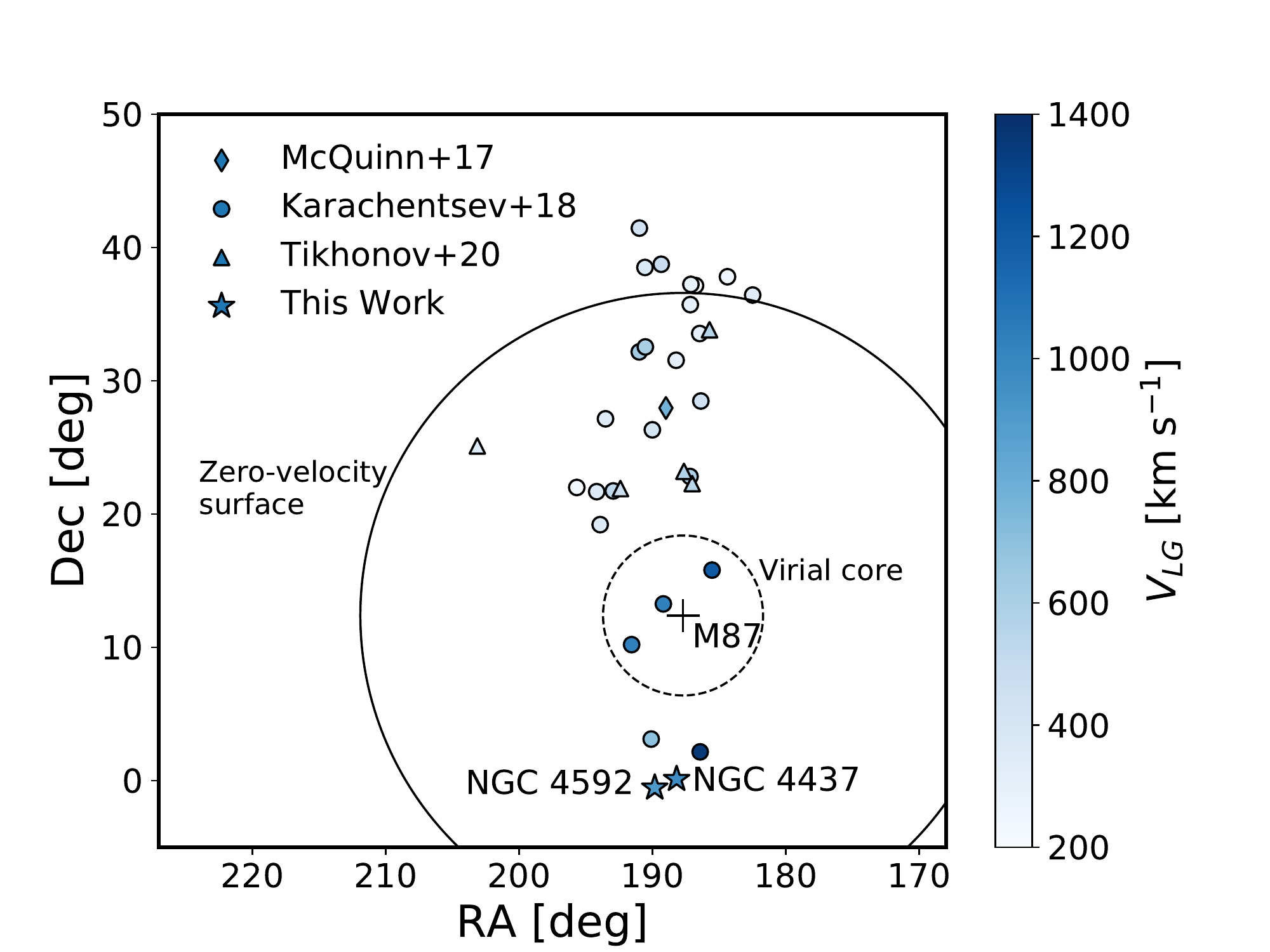} 
\caption{Map of the galaxies used for Virgo infall fitting.
Colors represent line-of-sight velocities with respect to the Local Group. 
Note that the corresponding velocity of the Virgo cluster is 988\kms~.
Lines show the radius of the zero-velocity surface (with a radius of 24\textdegree) and the virial radius (with a radius of 6\textdegree). 
}
\label{fig_map}
\end{figure}

\subsection{Virgo Infall Galaxy Sample}


\begin{deluxetable*}{lrrcrrrrrrrrr}[hbt!]
\rotate
\tabletypesize{\scriptsize}
\tablewidth{0pt} 

\caption{Virgo Infall Galaxy Sample }\label{tab-sample}
\tablehead{
\colhead{Name} 	& \colhead{R.A.\scriptsize{(J2000)}} & \colhead{Decl.\scriptsize{(J2000)}} & \colhead{Camera} & \colhead{$(F606W-F814W)_{{\rm TRGB}}$\tablenotemark{\footnotesize{a}}} & \colhead{$F814_{{\rm TRGB}}$\tablenotemark{\footnotesize{b}}} & \colhead{$D_{{\rm TRGB}}$\tablenotemark{\footnotesize{c}}} & \colhead{$A_{{\rm F814W}}$} & \colhead{$R_{{\rm VC}}$\tablenotemark{\footnotesize{d}}} & \colhead{$V_{{\rm LG}}$\tablenotemark{\footnotesize{e}}} & \colhead{$V_{{\rm VC, mi}}$} & \colhead{$V_{{\rm VC, ma}}$} & \colhead{$\lambda$} \\
\colhead{} 	& \colhead{[\textdegree]} & \colhead{[\textdegree]} & \colhead{} & \colhead{[mag]} & \colhead{[mag]} & \colhead{[Mpc]} & \colhead{[mag]} & \colhead{[Mpc]} & \colhead{[\kms]} & \colhead{[\kms]} & \colhead{[\kms]} & \colhead{[\textdegree]}
}
\startdata
UGC 07512  &  186.422208  &  2.158972  &  ACS  &  1.33 $^{+0.05}_{-0.02}$ (1) & 26.35 $\pm$ 0.08 (1) & 11.87 $\pm$ 0.53  &  0.03  &  5.26  &  1354 $\pm$ 6  &  --217  &  --461  &  146  \\ 
  VCC 2037  &  191.563750  &  10.205556  &  ACS  &  1.69 $^{+0.14}_{-0.01}$ (1) & 25.99 $^{+0.21}_{-0.19}$ (1) & 10.01 $^{+1.00}_{-0.91}$  &  0.04  &  6.56  &  1038 $\pm$ 1  &  --37  &  --54  &  169  \\ 
  IC 3583  &  189.181208  &  13.259331  &  ACS  &  1.79 $\pm$ 0.04 (1) & 26.00 $\pm$ 0.05 (1) & 9.92 $\pm$ 0.34  &  0.07  &  6.58  &  1034 $\pm$ 7  &  --44  &  --46  &  176  \\ 
  GR 34  &  185.531542  &  15.799111  &  ACS  &  1.69 $^{+0.11}_{-0.19}$ (1) & 25.91 $^{+0.29}_{-0.16}$ (1) & 9.64 $^{+1.31}_{-0.75}$  &  0.04  &  6.90  &  1206 $\pm$ 8  &  --205  &  --223  &  170  \\ 
  NGC 4437  &  188.189958  &  0.115028  &  ACS  &  1.23 $\pm$ 0.10 (3) & 25.82 $\pm$ 0.07 (6) & 9.28 $\pm$ 0.40  &  0.04  &  7.68  &  972 $\pm$ 5  &  90  &  --7  &  153  \\ 
  NGC 4600  &  190.095662  &  3.117750  &  ACS  &  1.27 $\pm$ 0.05 (1) & 25.78 $\pm$ 0.06 (1) & 9.07 $\pm$ 0.34  &  0.04  &  7.69  &  697 $\pm$ 2  &  318  &  297  &  159  \\ 
  NGC 4592  &  189.828067  &  -0.532008  &  WFC3  &  1.17 $\pm$ 0.06 (3) & 25.77 $\pm$ 0.03 (6) & 9.07 $\pm$ 0.27  &  0.03  &  7.93  &  912 $\pm$ 3  &  150  &  57  &  152  \\ 
  NGC 4559  &  188.990196  &  27.959992  &  ACS  &  1.17 $\pm$ 0.10 (4) & 25.72 $\pm$ 0.04 (5) & 8.87 $\pm$ 0.30  &  0.03  &  8.30  &  780 $\pm$ 2  &  287  &  203  &  148  \\ 
  NGC 4656  &  190.990542  &  32.168139  &  ACS  &  1.04 $\pm$ 0.01 (1) & 25.43 $\pm$ 0.04 (1) & 7.80 $\pm$ 0.25  &  0.02  &  9.55  &  644 $\pm$ 1  &  430  &  353  &  144  \\ 
  AGC 223231  &  185.719625  &  33.828528  &  ACS  &  0.97 $\pm$ 0.08 (4) & 25.38 $\pm$ 0.10 (4) & 7.64 $\pm$ 0.42  &  0.02  &  9.79  &  568 $\pm$ 1  &  500  &  447  &  142  \\ 
  AGC 229379  &  187.641675  &  23.205539  &  ACS  &  1.05 $\pm$ 0.11 (4) & 25.13 $\pm$ 0.10 (4) & 6.78 $\pm$ 0.37  &  0.03  &  9.92  &  571  &  438  &  421  &  162  \\ 
  NGC 4631  &  190.533375  &  32.541500  &  ACS  &  1.33 $\pm$ 0.01 (1) & 25.31 $\pm$ 0.02 (1) & 7.36 $\pm$ 0.20  &  0.03  &  9.93  &  604 $\pm$ 3  &  462  &  395  &  145  \\ 
  NGC 4455  &  187.183808  &  22.820447  &  ACS  &  1.13 $^{+0.02}_{-0.04}$ (1) & 25.06 $\pm$ 0.07 (1) & 6.54 $\pm$ 0.27  &  0.03  &  10.12  &  588 $\pm$ 1  &  420  &  402  &  163  \\ 
  AGC 742601  &  192.403542  &  21.917806  &  ACS  &  1.14 $\pm$ 0.09 (4) & 25.04 $\pm$ 0.09 (4) & 6.42 $\pm$ 0.32  &  0.05  &  10.25  &  472 $\pm$ 3  &  531  &  523  &  163  \\ 
  AGC 223254  &  187.020042  &  22.290167  &  ACS  &  1.16 $\pm$ 0.06 (4) & 25.01 $\pm$ 0.10 (4) & 6.39 $\pm$ 0.34  &  0.03  &  10.26  &  543 $\pm$ 4  &  461  &  448  &  164  \\ 
  IC 3840  &  192.942083  &  21.735194  &  ACS  &  1.05 $\pm$ 0.02 (1) & 24.85 $\pm$ 0.06 (1) & 5.86 $\pm$ 0.22  &  0.06  &  10.79  &  536 $\pm$ 3  &  469  &  454  &  163  \\ 
  KK 144  &  186.371458  &  28.482444  &  ACS  &  0.99 $\pm$ 0.02 (1) & 24.82 $\pm$ 0.06 (1) & 5.85 $\pm$ 0.22  &  0.04  &  11.00  &  453 $\pm$ 2  &  566  &  546  &  155  \\ 
  AGC 749241  &  190.007775  &  26.321183  &  ACS  &  0.99 $^{+0.02}_{-0.08}$ (4) & 24.66 $\pm$ 0.06 (3) & 5.47 $\pm$ 0.20  &  0.02  &  11.26  &  418 $\pm$ 1  &  591  &  579  &  159  \\ 
  Arp 211  &  189.340833  &  38.743889  &  ACS  &  1.06 $^{+0.04}_{-0.01}$ (1) & 24.87 $\pm$ 0.07 (1) & 6.02 $\pm$ 0.25  &  0.02  &  11.41  &  484 $\pm$ 2  &  590  &  523  &  140  \\ 
  AGC 238890  &  203.134583  &  25.114167  &  ACS  &  1.19 $\pm$ 0.07 (4) & 24.48 $\pm$ 0.12 (4) & 5.03 $\pm$ 0.32  &  0.02  &  11.89  &  356 $\pm$ 10  &  664  &  650  &  152  \\ 
  KDG 215  &  193.921000  &  19.209083  &  ACS  &  0.94 $\pm$ 0.02 (1) & 24.33 $\pm$ 0.07 (1) & 4.68 $\pm$ 0.19  &  0.03  &  11.90  &  362 $\pm$ 7  &  633  &  629  &  167  \\ 
  KK 177  &  195.674463  &  21.997394  &  ACS  &  1.10 $\pm$ 0.01 (2) & 24.39 $\pm$ 0.04 (2) & 4.75 $\pm$ 0.15  &  0.06  &  11.90  &  228 $\pm$ 2  &  767  &  772  &  163  \\ 
  DDO 133  &  188.226625  &  31.541111  &  WFPC2  &  1.05 $\pm$ 0.02 (1) & 24.43 $\pm$ 0.05 (1) & 4.90 $\pm$ 0.24  &  0.03  &  11.97  &  320 $\pm$ 1  &  694  &  688  &  153  \\ 
  NGC 4826  &  194.181838  &  21.682969  &  ACS  &  1.48 $\pm$ 0.02 (1) & 24.26 $\pm$ 0.03 (1) & 4.46 $\pm$ 0.13  &  0.06  &  12.15  &  365 $\pm$ 1  &  634  &  627  &  165  \\ 
  NGC 4395  &  186.453592  &  33.546928  &  ACS  &  1.30 $\pm$ 0.01 (1) & 24.36 $\pm$ 0.01 (1) & 4.75 $\pm$ 0.12  &  0.03  &  12.18  &  314 $\pm$ 1  &  705  &  696  &  151  \\ 
  DDO 126  &  186.771458  &  37.142583  &  WFPC2  &  1.01 $\pm$ 0.02 (1) & 24.44 $\pm$ 0.07 (1) & 4.93 $\pm$ 0.26  &  0.02  &  12.19  &  232 $\pm$ 2  &  783  &  808  &  145  \\ 
  PGC 038685  &  182.485292  &  36.434333  &  WFPC2  &  1.07 $^{+0.01}_{-0.03}$ (1) & 24.42 $\pm$ 0.13 (1) & 4.87 $\pm$ 0.36  &  0.04  &  12.23  &  341 $\pm$ 40  &  692  &  674  &  146  \\ 
  UGC 07605  &  187.161438  &  35.717481  &  WFPC2  &  0.96 $^{+0.03}_{-0.09}$ (1) & 24.33 $^{+0.13}_{-0.11}$ (1) & 4.69 $^{+0.35}_{-0.30}$  &  0.02  &  12.33  &  317 $\pm$ 1  &  708  &  696  &  148  \\ 
  DDO 127  &  187.119000  &  37.233639  &  WFPC2  &  0.97 $^{+0.01}_{-0.03}$ (1) & 24.35 $\pm$ 0.07 (1) & 4.72 $\pm$ 0.25  &  0.03  &  12.37  &  292 $\pm$ 5  &  734  &  730  &  146  \\ 
  LVJ 1243+4127  &  190.982083  &  41.456944  &  ACS  &  1.15 $\pm$ 0.02 (2) & 24.36 $\pm$ 0.09 (2) & 4.75 $\pm$ 0.23  &  0.03  &  12.56  &  444 $\pm$ 3  &  631  &  545  &  140  \\ 
  IC 3687  &  190.562917  &  38.503333  &  WFPC2  &  1.04 $\pm$ 0.02 (1) & 24.27 $\pm$ 0.08 (1) & 4.56 $\pm$ 0.25  &  0.03  &  12.56  &  381 $\pm$ 2  &  665  &  621  &  144  \\ 
  DDO 154  &  193.521875  &  27.149639  &  ACS  &  0.98 $\pm$ 0.02 (1) & 23.94 $\pm$ 0.03 (1) & 3.94 $\pm$ 0.11  &  0.01  &  12.75  &  354 $\pm$ 2  &  654  &  638  &  159  \\ 
  NGC 4244  &  184.373583  &  37.807111  &  ACS  &  1.21 $^{+0.03}_{-0.05}$ (1) & 24.13 $\pm$ 0.07 (1) & 4.26 $\pm$ 0.18  &  0.03  &  12.78  &  256 $\pm$ 1  &  765  &  765  &  146  \\ 
  \enddata
\tablenotetext{a~}{TRGB $(F606W - F814W)_{ACS}$ colors. References: (1) EDD; (2) \citet{kar18}; (3) this study; (4) visually measured from CMDs. 
}
\tablenotetext{b~}{TRGB $F814W_{{\rm ACS}}$ magnitudes. References: (1) EDD; (2) \citet{kar18}; (3) \citet{mcq14}; (4) \citet{tik20}; (5) \citet{mcq17}; (6) this study. 
}
\tablenotetext{c~}{TRGB distances calculated applying the \citet{jan20} calibration.}
\tablenotetext{d~}{Calculated for $D_C =$ 16.5 Mpc}
\tablenotetext{e~}{Local Group rest-frame velocities from NED. For AGC 229379, adopted from \citet{tik20}.}
\end{deluxetable*}


Combining our results for NGC 4437 and NGC 4592 with those in the literature, we prepare a list of galaxies located between the Virgo cluster and the Local Group for which TRGB distances are available.
We mainly use the sample of Virgo infall galaxies from \citet{kar18} and add several galaxies from the literature that satisfy the selection criterion presented in \citet{kar18}.
As will be discussed in Section \ref{sec_virgo_infall_fit}, an ambiguity is introduced when transforming observed velocities into Virgocentric velocities.
This ambiguity is a function of $\lambda$, which is defined as the angle between the line of sight toward a galaxy and the line connecting the galaxy with the cluster center.
If a galaxy is located at the line-of-sight direction toward the Virgo cluster ($\lambda =$ 180\textdegree), the Virgocentric velocity equals the line-of-sight velocity.
However, the ambiguity increases as $\lambda$ decreases. 
Therefore, \citet{kar18} adopted a selection criterion of $\lambda =$ [135\textdegree, 180\textdegree].

In this study, we adopt $\lambda =$ [140\textdegree, 180\textdegree] in order to further reduce ambiguity.
We adopt 25 galaxies from \citet{kar18}, excluding two galaxies with $\lambda =$ [135\textdegree, 140\textdegree].
In this sample, TRGB magnitudes of 22 galaxies were obtained from the Extragalactic Distance Database (EDD)\citep{jac09}\footnote{http://edd.ifa.hawaii.edu/}, two galaxies (KK 177 and LVJ 1243+4127) are from \citet{kar18}, and one galaxy (AGC 749241) is from \citet{mcq14}.
In addition, we find that five galaxies from \citet{tik20} satisfy $\lambda =$ [140\textdegree, 180\textdegree] and show Virgo infall motions: AGC 223231, AGC 223254, AGC 229379, AGC 238890, and AGC 742601.
Moreover, NGC 4559 for which the TRGB magnitude is measured by \citet{mcq17} also satisfies the criteria.
With our own  measurements for NGC 4437 and NGC 4592, we compile TRGB magnitudes of 33 galaxies in total.

Table \ref{tab-sample} shows a list of our sample galaxies.
All the magnitudes in the table are in the HST/ACS filter system.
The Johnson-Cousins $V$, $I$ magnitudes in \citet{tik20} sample and the magnitudes of six galaxies observed by WFPC2 in EDD sample are transformed to $F606W_{{\rm ACS}}$ and $F814W_{{\rm ACS}}$ magnitudes using synthetic transformations obtained by \citet{sir05}.
For NGC 4592, the $(F555W - F814W)_{{\rm WFC3}}$ color is converted to the $(F555W - F814W)_{{\rm ACS}}$ color using the transformation in \citet{jan15}, then converted to the $(F606W - F814W)_{{\rm ACS}}$ color applying Eq. (1) in \citet{jan17}. 
TRGB distances given in the table are derived using the \citet{jan20} calibration in order to maintain consistency.
In calculating the error budget of TRGB distances, we consider the uncertainties of TRGB magnitudes provided in the literature, \citet{jan20} calibration error 0.055 mag, the magnitude transformation error 0.02 mag from Johnson-Cousins $I$-band to ACS for the \citet{tik20} sample \citep{sir05}, the zero-point error 0.07 mag of WFPC2 camera \citep{freed01}, and the galactic extinction error (half of the extinction\footnote{The foreground extinction corrections for the sample galaxies are reasonably small (A$_{F814W} <$ 0.05), as listed in Table 3. We adopt a conservative estimate of errors, taking half of A$_{F814W}$, considering a systematic uncertainty of the extinction map (e.g., $\sigma_{E(B-V)} \approx$ 0.03 in
\citet{schlegel98}).}).
Radial velocities in the Local Group frame were obtained from NASA/IPAC Extragalacitc Database (NED)\footnote{https://ned.ipac.caltech.edu/}.

Spatial distributions of the galaxies used in this study are shown in Figure \ref{fig_map}.
The cross mark denotes the location of M87, the approximate dynamical center of the Virgo cluster.
The radius of the virial core ($R \sim$ 6\textdegree \citep{hof80}) of the Virgo cluster 
is marked as the dashed line circle and the zero-velocity surface ($R \sim$ 24\textdegree) as the solid line circle.
Colors represent radial velocities in the Local Group frame.
Galaxies are mainly distributed along the north-south direction.
It is noted that NGC 4437 and NGC 4592 are rare infall examples in the southern region of the Virgo cluster.
Galaxies closer to M87 tend to have larger radial velocities than those farther from M87, showing infall motion toward the Virgo center.

\subsection{Theoretical Description of the Virgo Infall Model}

\citet{pei06, pei08} derived the velocity-distance relations of a spherical shell of radius $R$ moving radially in the gravitational field created by a mass $M$ inside a shell, including the cosmological constant term.
This is applicable to regions outside the virial core, where the assumption of purely radial velocities can be held. 
They numerically solved an equation of motion of a shell
\begin{equation}\label{eq:eom}
    {d^2R \over dt^2} = - {GM \over R^2} + \Omega_\Lambda H_0^2 R
\end{equation} 
to obtain the velocity and distance of the present age of the universe, assuming $\Omega_\Lambda =$ 0.7.

We followed the same method but using $\Omega_\Lambda =$ 0.685 in  \citet{Planck18} instead of 0.7.
Dimensionless variables $x = t H_0$, $y = R/R_0$, and $u = \dot{R}/H_0 R_0$ are defined for calculation, where $R_0$ is the present radius of the zero-velocity surface.
At the radius of zero-velocity surface, where the relative velocity with respect to the Virgo center is zero, the cluster is separated from the Hubble expansion.
In addition, introducing the constant $A = 2GM/(H_0^2 R_0^3)$, Eq. \eqref{eq:eom} can be rewritten as:
\begin{equation}\label{eq:eom_nondim}
    {d^2y \over dx^2} = - {A \over 2y^2} + \Omega_\Lambda y
\end{equation}
The value of $A$ is determined from boundary conditions (see \citet{pei08} for details), $A =$ 3.7094 for $\Omega_\Lambda =$ 0.685, yielding the expression of mass of the Virgo cluster within the zero-velocity surface as:
\begin{equation} \label{eq:mass}
    M = {1.855 \over G} \times H_0^2 R_0^3 = 4.29 \times 10^{12} h^2 { ( {R_0 \over [{\rm Mpc}]} )}^3 M_\odot,
\end{equation}
\noindent where $h$ is the local Hubble constant in units of 100 \kmsMpc~ and $R_0$ is the present radius of the zero-velocity surface in Mpc.
Present velocity and distance solutions of Eq. \eqref{eq:eom_nondim} with various initial energies are well described by the form $u = -b/y^n + by$, where $b =$ 1.335 and $n =$ 0.702. The values of these parameters are similar to those given by \citet{pei06, pei08}. 

From these results we derive a velocity-distance relation:
\begin{equation}\label{eq:infall}
    v = f(R) = - {0.940 H_0 \over R^n}({GM \over H_0^2})^{(n+1)/3} + 1.335 H_0 R. 
\end{equation}
\noindent Once we have cluster-centric distances and velocities of galaxies around a cluster, we can determine $H_0$ and $R_0$ (or $M$) using this equation.

\begin{figure*} 
\centering
\includegraphics[scale=0.75]{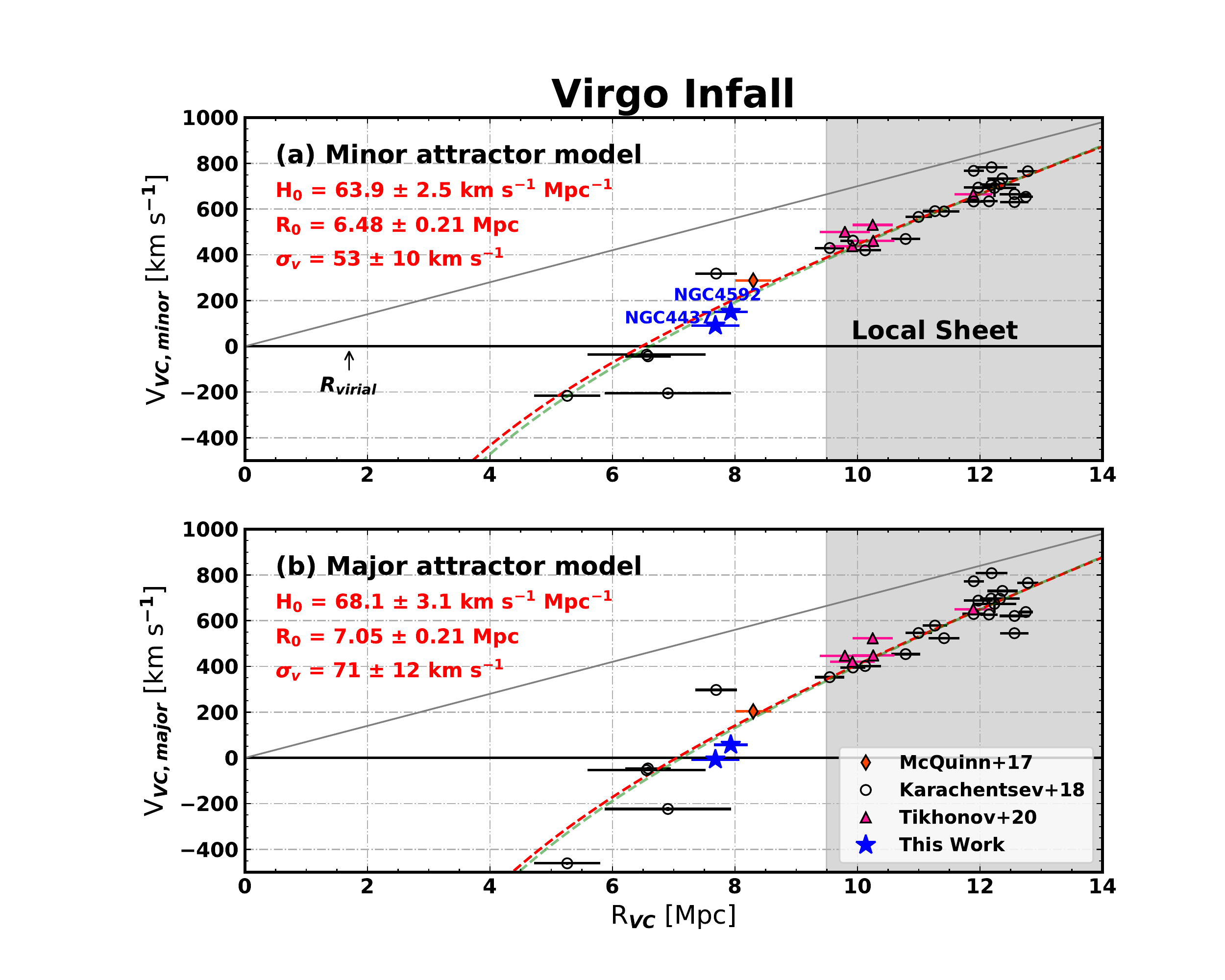} 
\caption{The Hubble diagram of the sample galaxies in the Virgo cluster frame:
Virgocentric velocity vs. Virgocentric distance.
Radial velocities are transformed into Virgocentric velocities using 
(a) the minor attractor model and 
(b) the major attractor model.
NGC 4437 and NGC 4592, of which TRGB distances are measured in this study, are marked as blue starlets. 
Fitted models using the MCMC method and using the least-squares method are marked as red and green dashed lines, respectively.
The gray straight lines represent the unperturbed Hubble flow for $H_0=70$ \kmsMpc.
}

\label{fig_vel-disp}
\end{figure*}


\begin{figure} 
\centering
\includegraphics[scale=0.6]{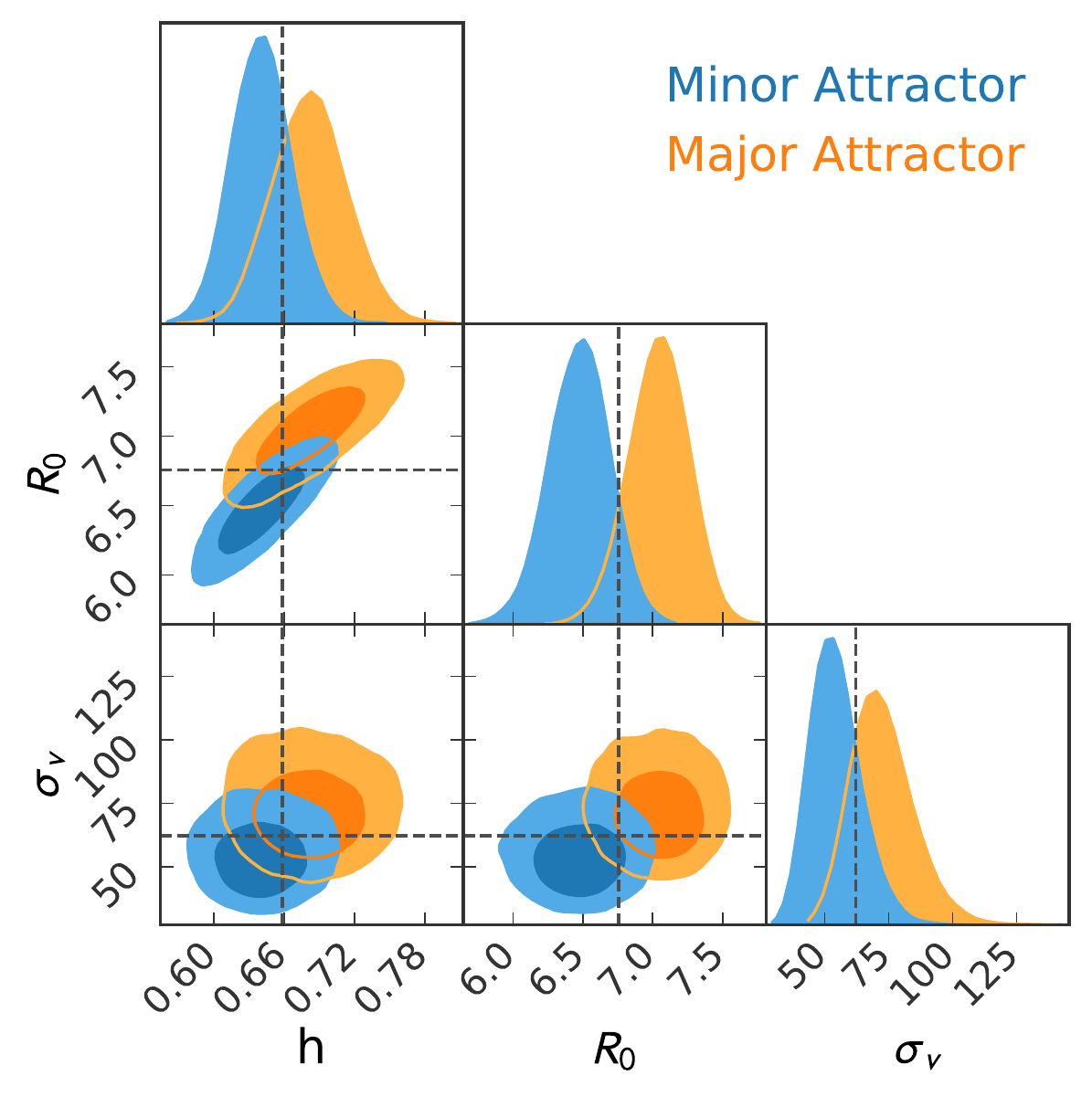} 
\caption{Posterior probability distributions for the parameters ($h$ [100\kmsMpc], $R_0$ [Mpc], $\sigma_v$ [\kms]) estimated by minor (blue) and major (orange) attractor models. Two contour levels indicate 68\% and 95\% levels, respectively.  Estimated parameter values and their errors are obtained from the median (dashed lines) and standard deviation of the sum of two posterior probability distributions.}
\label{fig_mcmc}
\end{figure}

\subsection{Fitting the Virgo Infall Pattern}\label{sec_virgo_infall_fit}

In order to fit our data to the model in Equation \eqref{eq:infall}, we first need to transform the measured distances and velocities of galaxies into Virgocentric distances and Virgocentric velocities.
To determine the distance and velocity of the Virgo cluster, we use the catalog of galaxies around M87 \citep{kas20}.
We select 78 early-type galaxies that are located within the angular virial radius ($\theta < 6$\textdegree) and in the distance range of $12<D[{\rm Mpc}]<22$, derived from the surface brightness fluctuation (SBF) method.
The distance distribution of the selected galaxies shows a significant concentration at $D \sim 16.5$ Mpc.
From these galaxies we derive a mean distance 
$<D_C> =$ 16.5 $\pm$ 0.1 Mpc and a mean velocity $<V_C>_{LG} =$ 988 $\pm$ 61 \kms, which are adopted as the distance and velocity of Virgo for the following analysis. 
These values are similar to those in \citet{kas18b} and \citet{kar18}.

We calculate Virgocentric distances as $R_{VC}^2 = D_g^2 + D_C^2 - 2D_g \times D_C \times cos\Theta$, where $\Theta$ is the angular separation of a galaxy from the Virgo center (M87) and $D_g$ is the TRGB distance of the galaxy.
Due to our ignorance of tangential velocities of the galaxies,  
radial velocities are transformed into Virgocentric velocities assuming two extreme cases of the mass of the overdensity \citep{kar10,kas18}.
One is a minor attractor model where the peculiar velocities of satellite galaxies around a group or cluster are much smaller than the velocities of regular Hubble flow (see Figure 3 in \citet{kas18}).
\begin{equation}
    V_{VC, {\rm minor}} = V_g \times cos\lambda - V_c \times cos(\lambda + \Theta)
\end{equation}
where $V_g$ is the radial velocity of the galaxy in the Local group rest frame.
The other is a major attractor model where satellite galaxies show significant infall motion toward the center of a galaxy group or cluster.
\begin{equation}
    V_{VC, {\rm major}} = [V_g - V_C \times cos\Theta] / cos\lambda
\end{equation}
The details of each model are described in \citet{kar10} and \citet{kas18}. 
Note that the difference between $V_{VC, {\rm minor}}$ and $V_{VC, {\rm major}}$ increases as $\lambda$ deviates from 180\textdegree: galaxies with $\lambda =$ [140\textdegree, 160\textdegree] have an average difference of 50 \kms~ whereas those with $\lambda =$ [160\textdegree, 180\textdegree] have 10 \kms.
True Virgocentric velocities are expected to lie between velocities transformed by the minor attractor model and the major attractor model. 

Virgocentric distances and Virgocentric velocities of our galaxy sample are shown in Figure \ref{fig_vel-disp}, (a) assuming the minor attractor model and (b) assuming the major attractor model.
Compared with the unperturbed Hubble flow (gray lines), our sample galaxies show infall motions toward the Virgo cluster.
$V_{VC, {\rm major}}$ values are 
generally smaller than $V_{VC, {\rm minor}}$ values for each galaxy.
Note that the velocity scatter of galaxies in the Local Sheet ($D_{g} <$ 7 Mpc \citep{tul08}, gray shaded region ($R_{VC} >$ 9.5 Mpc)) is smaller than that of galaxies closer to the Virgo cluster.

Then we fit the Virgocentric distances and Virgocentric velocities of galaxies using the Bayesian approach.
The likelihood of observing a galaxy at Virgocentric distance $R_{i}$ given parameters $H_0$, $R_0$, and error $\sigma$ is modeled as
\begin{equation} \label{eq:likelihood}
  p(v_i|R_i, H_0, R_0, \sigma) = {1 \over \sqrt{2\pi\sigma^{2}}} \exp{\left(-{{(f(R) - v_i)}^2 \over 2\sigma^2}\right)}
\end{equation}
where
\begin{equation} \label{eq:sigma}
\sigma^2 = err(v_i)^2 + err(R_i)^2\times ({{\partial f} \over {\partial R}})^2 +\sigma_v^2.
\end{equation}
Here, $err(v)$ and $err(R)$ are measurement errors of velocity and distance.
$\sigma_v$ is an intrinsic velocity dispersion independent of measurement errors, which can be interpreted as a scatter due to peculiar motions.
${{\partial f} / {\partial R}}$ denotes the gradient of the model Hubble flow at given distance and parameters.
Then the posterior probability distribution is given as
\begin{equation} \label{eq:posterior}
  p(H_0, R_0, \sigma_v|v) \propto \displaystyle\prod_{i} p(v_i|H_0, R_0, \sigma_v) \times p(H_0, R_0, \sigma_v)
\end{equation}
Here, we assume a uniform prior probability distribution ($p(H_0, R_0, \sigma_v)$) over 40 \kmsMpc $<$ $H_0$ $<$ 90 \kmsMpc, 3 Mpc $<$ $R_0$ $<$ 9 Mpc, and 10 \kms $<$ $\sigma_v$ $<$ 200 \kms.

We use \texttt{emcee} \citep{fore13}, a python implementation of the ensemble sampler for MCMC. 
Values of the local Hubble constant $H_0$, the radius of zero-velocity surface $R_0$, and the intrinsic velocity dispersion $\sigma_v$ are estimated from both the minor and major attractor models.

We also tried the least-squares fitting method for the comparison with \citet{kar18}.
\citet{kar18} used the least-squares fitting method for both the minor and major attractor models, and averaged the results to get mean values.
We follow the same fitting method, but additionally conducted Monte Carlo simulations in order to estimate fitting errors.
First, each data point ($R_{VC}$, $V_{VC}$) is shifted randomly 5000 times assuming a Gaussian distribution with standard deviation of measurement errors. Then $H_0$ and $R_0$ parameters are obtained 5000 times using the least-squares method.
Uncertainties of the parameters are obtained from standard deviations of 5000 trials.
Note that the intrinsic velocity dispersion $\sigma_v$ is unable to be modeled using the least-squares method.

\subsection{Fitting Results}\label{sec_virgo_infall_fit_results}

The top panel in Table \ref{tab_result_mcmc} lists the results of the MCMC fitting.
Using $V_{VC, {\rm minor}}$ for $v$, we obtain $H_{0, {\rm minor}} =$ 63.9 $\pm$ 2.5 \kmsMpc~, $R_{0, {\rm minor}} =$ 6.48 $\pm$ 0.21 Mpc, and $\sigma_{v, {\rm minor}} =$ 53 $\pm$ 10 \kms.
In addition, using $V_{VC, {\rm major}}$ for $v$, we get $H_{0, {\rm major}} =$ 68.1 $\pm$ 3.1 \kmsMpc~, $R_{0, {\rm major}} =$ 7.05 $\pm$ 0.21 Mpc, and $\sigma_{v, {\rm major}} =$ 71 $\pm$ 12 \kms.
The rms errors of the fit are 73 \kms~ and 84 \kms~, respectively. 
All the parameters $H_0$, $R_0$, and $\sigma_v$ are slightly larger for the major attractor model than those of the minor attractor model.
The fitted lines are shown as red dashed lines in Figure \ref{fig_vel-disp}.
Figure \ref{fig_mcmc} displays the posterior probability distributions of the parameters for both minor and major attractor models.

The bottom panel in Table \ref{tab_result_mcmc} and the green dashed lines in Figure \ref{fig_vel-disp} show the results of the least-squares fitting.
The values of the parameters derived using this method are slightly ($\sim$ 1.5\%) larger than those obtained using the MCMC method, but they are consistent within the error range.

Since the minor and major attractor models are two extreme cases, it is likely that the true values lie between parameter values estimated by the two models.
We sum probability distributions obtained from two cases (histograms in Figure \ref{fig_mcmc}),
and derive median values and standard errors from these. 
In doing so, errors of the parameters are conservatively estimated.
We obtain average parameter values as $H_0 =$ 65.8 $\pm$ 3.5 \kmsMpc, $R_0 =$ 6.76 $\pm$ 0.35 Mpc, and $\sigma_v =$ 62 $\pm$ 14 \kms (right column of Table \ref{tab_result_mcmc}).
The implications of $H_0$ will be discussed in Section \ref{sec_H0}.
Here, we briefly discuss our results of $R_0$, rms errors of the fit, and $\sigma_v$.

We calculate the mass of the Virgo cluster within $R_0$ with Equation \eqref{eq:mass}, obtaining $M(R_0) =$ (5.7 $\pm$ 1.5) $\times 10^{14} M_{\odot}$.
This value is very similar to the virial mass estimate, $M_{vir} =$ (6.3 $\pm$ 0.9) $\times 10^{14} M_{\odot}$ for the virial radius of 1.7 Mpc, derived from  velocity dispersion of all types of galaxies by \citet{kas20}.
Assuming that only early-type galaxies are in the relaxed state and adopting the velocity dispersion of these galaxies,
they derived a slightly smaller value, $M_{vir} =$ 4.1 $\times 10^{14} M_{\odot}$.
Our estimation of $M(R_0)$ lies between the two values, suggesting that $M(R_0) \sim M_{vir}$.
This confirms that the outer region of the Virgo cluster core does not contain a considerable amount of dark matter, as suggested by previous studies \citep{kar14, kar18, kas18b, kas20}.

Among the previous studies that estimated $R_0$, \citet{kar18} fitted TRGB distances and velocities of the galaxies in front of the Virgo cluster to the velocity-distance relation but they used a slightly different form from ours, which will be described in Section 5.3.
They obtained $R_0 =$ 7.3 $\pm$ 0.3 Mpc and the mass of the Virgo cluster within $R_0$ as $M(R_0) =$ (7.64 $\pm$ 0.91) $\times 10^{14} M_{\odot}$ where Planck model parameters \citep{Planck18} are used.
Our results are slightly smaller than those of \citet{kar18}.
This is mainly because  the fitting methods and the adopted distances of the Virgo cluster are different in the two studies.
The value of $R_0$ obtained by the least-squares method in this study is slightly closer to the value of $R_0$ in \citet{kar18}.
In addition, the adopted distance of the Virgo cluster in \citet{kar18} is 16.65 Mpc, 0.15 Mpc larger than the value in this study, accounting for the larger value of $R_0$.
The error in our study is larger than the error given by \citet{kar18}. 
This is because we include the error of both $H_0$ and $R_0$, while \citet{kar18} include only the error of $R_0$.

We compare the rms velocity errors (or observed velocity dispersions) in this study (73 \kms~ and 84 \kms~ for the minor and major attractor models) to those in the previous studies.
\citet{pei08} obtained the velocity dispersion as 345 \kms~ by fitting velocities and Tully-Fisher distances of 27 galaxies in the Virgo infall region. 
This value is much larger than those in this study.
The large uncertainties in Tully-Fisher distances might have attributed to their large values.
In fact, \citet{kar18} suggested a much smaller velocity dispersion by using only the galaxies with TRGB distances to fit the Virgo infall pattern.
They obtained the velocity dispersion to be 92 \kms~ and 105 \kms~ for the minor and major attractor models,  mentioning that the Hubble flow around the Virgo cluster is cold.
Our velocity dispersion values based on the larger sample are even smaller (by about 20\%) than those of \citet{kar18}, supporting the coldness of the Hubble flow. 

The intrinsic velocity dispersion, $\sigma_v$, is a useful quantity, which can be interpreted as the velocity dispersion minus the dispersion contributed by measurement errors.
We obtain $\sigma_v =$ 62 $\pm$ 14 \kms, which is about 80\% of the rms errors. 
This shows that the Hubble flow around Virgo is even colder than suggested in the previous studies \citep{kar18}: 78 \kms~ and 90 \kms~ for the minor and major attractor models.
The value for Virgo is about 1.6 times larger than the $\sigma_v$ derived from the Local Group Hubble flow, $\sigma_v =$ 38 \kms \citep{pen14}.

\begin{deluxetable*}{lccc}
\tabletypesize{\footnotesize}
\caption{Virgo Infall Fitting Results}\label{tab_result_mcmc}
\tablewidth{0pt}
\tablehead{\colhead{} & \colhead{Minor Attractor}& \colhead{Major Attractor} & \colhead{Average}}
\startdata
MCMC & & \\ \hline 
$H_{0}$ [\kmsMpc] & 63.9 $\pm$ 2.5 &68.1 $\pm$ 3.1 &65.8 $\pm$ 3.5 \\ 
$R_{0}$ [Mpc] & 6.48 $\pm$ 0.21 & 7.05 $\pm$ 0.21 & 6.76 $\pm$ 0.35 \\ 
$\sigma_{v}$ [\kms] & 53 $\pm$ 10 & 71 $\pm$ 12 & 62 $\pm$ 14 \\ 
RMS [\kms] & 72.9 & 83.6 & \\ 
Virgo Mass [M$_\odot$] & (4.8 $\pm$ 0.8) $\times$ 10$^{14}$& (7.0 $\pm$ 1.3) $\times$ 10$^{14}$& (5.7 $\pm$ 1.5) $\times$ 10$^{14}$ \\ \hline  
Least-Squares + Monte-Carlo & & \\ \hline 
$H_{0}$ [\kmsMpc] & 65.1 $\pm$ 2.4 &68.9 $\pm$ 2.8 &66.8 $\pm$ 3.2 \\ 
$R_{0}$ [Mpc] & 6.62 $\pm$ 0.22 & 7.13 $\pm$ 0.21 & 6.86 $\pm$ 0.33 \\ 
RMS [\kms] & 71.2 & 82.6 & \\ 
Virgo Mass [M$_\odot$] & (5.3 $\pm$ 0.9) $\times$ 10$^{14}$& (7.4 $\pm$ 1.3) $\times$ 10$^{14}$& (6.2 $\pm$ 1.4) $\times$ 10$^{14}$ \\ 
\enddata
\end{deluxetable*}


\section{Discussion}

\subsection{The NGC 4437 Group}\label{sec_group}

Two previous studies classified NGC 4437 and NGC 4592 as a galaxy group, but the separation of the two galaxies was considered large, lacking grounds for their membership \citep{tul13,kar&na13}.
\citet{tul13} grouped NGC 4437 with NGC 4592 and another galaxy NGC 4544 using redshift information, and presented their weighted average of existing distance measurements to be 8.58, 11.69, and 18.11 Mpc respectively. 
Given the faintness of NGC 4437 ($M_{B_T} =$ --18.7 mag \citep{dev91}, about an order of magnitude fainter than M31), their separations are too large for them to be considered as a group.
\citet{kar&na13} studied galaxies in the southern part of the Virgo cluster and identified NGC 4437, NGC 4592, and CGCG 014-054 as a foreground galaxy group. 
They adopted their Tully-Fisher distances as 9.7, 11.6, and 9.6 Mpc, respectively. However, the separations between NGC 4592 and other galaxies are too large for them to be considered as a group. 

As presented in Section \ref{sec_TRGB}, our distances to NGC 4437 and NGC 4592 are very similar, showing that they are spatially adjacent.
Their three-dimensional spatial separation is only 0.29 Mpc and its 1$\sigma$ range is [0.26, 0.59] Mpc.
We conclude that they are indeed a physical pair located close to the zero-velocity surface of the Virgo cluster ($R_0 =$ 6.76 $\pm$ 0.35 Mpc).
In fact, we measured SBF distances to dwarf galaxies found in a wide HSC 5\textdegree $\times$ 5\textdegree~ image 
centered on NGC 4437 and found five dwarf galaxies to be located at distances similar to NGC 4437 and NGC 4592, implying that they are probable members of the NGC 4437 group.
These results will be presented in the future paper (Kim et al. (2020) in preparation).

\subsection{Systematic Uncertainties of $H_0$}\label{sec_err}

There are at least four kinds of systematic uncertainties present in our method for obtaining $H_0$: (1) uncertainties from velocity-distance models, (2) uncertainties due to ignorance of tangential velocities, (3) zero-point uncertainties in the TRGB measurement, and (4) uncertainties in the distance and velocity of the Virgo cluster. 
In this Section, we describe possible sources of systematic uncertainties in our $H_0$ determination.

First, our method is model-dependent.
The velocity-distance model assumes a spherically symmetric mass distribution of the Virgo cluster, which is a simplified model.
However, we conjecture that the uncertainties arising from the imperfect model are not significant.
\citet{pen14} carried $N$-body experiments of the Hubble flow perturbed by the Local Group, using two mass distribution models: central point mass and a pair of point masses (corresponding to the Milky Way and M31).
They determined the mass of the Local Group and cosmological constants from synthetic data generated with both models.
Although different mass distribution models resulted in different values of the Local Group mass, the values of the Hubble constant remained unchanged.
This implies that the possible asymmetric mass distribution in the core of the Virgo cluster might not influence the value of $H_0$ significantly.
Moreover, since there are neither conspicuous large-scale structures nor massive galaxy groups between the Local Group and the Virgo cluster, it is likely that the Hubble constant is not biased due to mass distributions in the outskirts.

Nevertheless, the model fitting range may introduce systematic uncertainties. 
The model should be applied to the galaxies that have been under the influence of gravitational force exerted by the Virgo mass $M$ from the early universe.
That is, galaxies in the outer region might not follow the model.
Thus, it is crucial to restrict the fitting range of distances.
In fact, \citet{nas11} fitted the velocity-distance relation using velocities and distances of galaxies in front of the Fornax cluster, which are located at $R_{{\rm Fornax}} < R_{{\rm max}}$ where $R_{{\rm max}} H_0 = v(R_{{\rm max}})$.
$R_{{\rm max}}$ is approximately 2.2 times the radius of zero-velocity surface. 
By this criterion, they separated infalling galaxies that follow the velocity-distance relation (Eq. \eqref{eq:infall}) and other galaxies that follow Hubble expansion only.
All of our 33 galaxies are located at $R_{VC} < 1.9 R_0$, so it can be considered reasonable to assume that they have been in the infall region from the early universe.
Still, the true border is ambiguous and we cannot exclude a possibility that the criterion could be modified.

Second, we conjectured that the true $H_0$ is likely to lie in between $H_{0, {\rm minor}}$ and $H_{0, {\rm major}}$.
Since 
tangential velocities of the galaxies are not available, radial velocities are transformed into Virgocentric velocities assuming two extreme cases.
$H_0$ is diminished to 63.9 $\pm$ 2.5 \kmsMpc~ when the minor attractor model is true, and $H_0$ is increased to 68.1 $\pm$ 3.1 \kmsMpc~ when the major attractor model is true.
The uncertainty due to ambiguous velocity transformation is reflected into the error range ($\pm$ 3.5 \kmsMpc) we derived in this study. 

One method of reducing systematic uncertainties arising from the different velocity transformation models is using only the galaxies with high $\lambda$.
In Table \ref{tab_result_lambdif} we present $H_0$ values obtained from different constraints on $\lambda$.
As expected, using more stringent ranges for $\lambda$ results in the gaps between $H_0$s of the minor attractor model and the major attractor model decreasing, but shows larger random errors due to smaller sample size.
Using $\lambda >$ 145\textdegree~ or $\lambda >$ 150\textdegree, $H_0$s increase both for the minor and major attractor models, but not significantly considering random errors.
The standard deviation of four $H_0$s using different ranges of $\lambda$ is $\sim$2.0 \kmsMpc.
Although systematic uncertainty decreases when using a smaller number of galaxies with larger $\lambda$, the increase of random error becomes significant.
Thus we adopt $\lambda >$ 140\textdegree~ for our final results.

\begin{deluxetable}{lcccc}
\tabletypesize{\footnotesize}
\caption{$H_0$ [\kmsMpc] for Different Ranges of $\lambda$ }\label{tab_result_lambdif}
\tablewidth{0pt}
\tablehead{\colhead{$\lambda$ range} & \colhead{N} & \colhead{Minor Attractor}& \colhead{Major Attractor} & \colhead{Average}}
\startdata
$\lambda >$ 140\textdegree & 33 & 63.9 $\pm$ 2.5 &68.1 $\pm$ 3.1 &65.8 $\pm$ 3.5 \\ 
$\lambda >$ 145\textdegree & 27 & 65.8 $\pm$ 2.6 &70.6 $\pm$ 3.1 &68.1 $\pm$ 3.7 \\ 
$\lambda >$ 150\textdegree & 20 & 65.3 $\pm$ 3.6 &67.8 $\pm$ 4.2 &66.4 $\pm$ 4.1 \\ 
$\lambda >$ 155\textdegree & 15 & 63.3 $\pm$ 4.8 &63.2 $\pm$ 5.0 &63.2 $\pm$ 4.9 \\ 
\enddata
\end{deluxetable}

Third, our $H_0$ determination is subject to uncertainties of TRGB calibration.
Quadratic sum of the statistical uncertainty and the systematic uncertainty in the TRGB calibration of \citet{jan20} is 0.055 mag.
This results in $\pm$ 1.1 \kmsMpc~ uncertainty in $H_0$. 
We also test $H_0$ determination using different calibrations.
Using the calibration of \citet{freed20} anchored on the distance to the Large Magellanic Cloud, we obtain $H_{0, {\rm minor}} =$ 64.0 $\pm$ 2.5 \kmsMpc~ and $H_{0, {\rm major}} =$ 68.4 $\pm$ 3.1 \kmsMpc~, yielding $H_0 =$ 66.0 $\pm$ 3.5 \kmsMpc.

In addition, given that TRGB colors of a few galaxies in our sample marginally belong to the blue, constant TRGB magnitude range, we use calibrations by \citet{riz07} and \citet{jan17} as well.
\citet{riz07} obtained a linear relation between TRGB colors and F814W TRGB magnitudes.
Using their calibration, we obtain $H_{0, {\rm minor}} =$ 66.6 $\pm$ 2.7 \kmsMpc~ and $H_{0, {\rm major}} =$ 71.6 $\pm$ 3.0 \kmsMpc~, yielding $H_0 =$ 69.0 $\pm$ 3.8 \kmsMpc.
Moreover, \citet{jan17} presented a quadratic relation between RGB colors and TRGB magnitudes.
Adopting their calibration using NGC 4258 as a distance anchor with the zero-point replaced with the updated value from \citet{jan20}, we get $H_{0, {\rm minor}} =$ 64.2 $\pm$ 2.5 \kmsMpc~ and $H_{0, {\rm major}} =$ 68.3 $\pm$ 3.1 \kmsMpc~, resulting in $H_0 =$ 66.1 $\pm$ 3.5 \kmsMpc.
Thus, all these values from different calibrations agree within errors.

Lastly, the $H_0$ determination depends on the distance and velocity of the Virgo cluster.
Adopting the values as $<D_C> =$ 16.5 $\pm$ 0.1 Mpc and $<V_{C}>_{LG} =$ 988 $\pm$ 61 \kms~, and assuming Gaussian error distributions, we calculate systematic errors using the Monte Carlo approach.
The smaller the $D_C$, the larger the $H_0$: $\pm$ 0.1 Mpc uncertainty of $D_C$ results in $\mp$ 0.2 \kmsMpc~ uncertainty in $H_0$.
In addition, the larger the $V_C$, the larger the $H_0$: $\pm$ 61 \kms~ uncertainty of $V_C$ results in $\pm$ 2.1 \kmsMpc~ uncertainty in $H_0$.
Thus, combined with the calibration uncertainty ($\pm$ 1.1 \kmsMpc) described above, the total systematic uncertainty is 2.4 \kmsMpc~, resulting in $H_0 =$ 65.8 $\pm$ 3.5($stat$) $\pm$ 2.4($sys$) \kmsMpc.
The calculated error budgets are summarized in Table \ref{tab_result_syserr}.

\begin{deluxetable}{lccc}
\tabletypesize{\footnotesize}
\caption{Error Budget of $H_0$ [\kmsMpc] }\label{tab_result_syserr}
\tablewidth{0pt}
\tablehead{\colhead{Parameter} & \colhead{Minor Attractor} & \colhead{Major Attractor} & \colhead{Average}}
\startdata
Random (Fitting) & 2.5 & 3.1 & 3.5\tablenotemark{\footnotesize{a}} \\ \hline \hline
Systematic & & & \\ \hline
TRGB Calibration & 1.2 & 1.0 & 1.1 \\ 
$<D_C>$ & 0.1 & 0.2 & 0.2 \\
$<V_C>_{LG}$ & 2.1 & 2.2 & 2.1 \\ \hline
Systematic Total & 2.5 & 2.4 & 2.4 \\
\enddata
\tablenotetext{a~}{
The uncertainty due to the ambiguous velocity transformation is reflected into the error range of the average value.
}
\end{deluxetable}

\subsection{The Local $H_0$}\label{sec_H0}


\begin{figure*}
\centering
\includegraphics[scale=0.8]{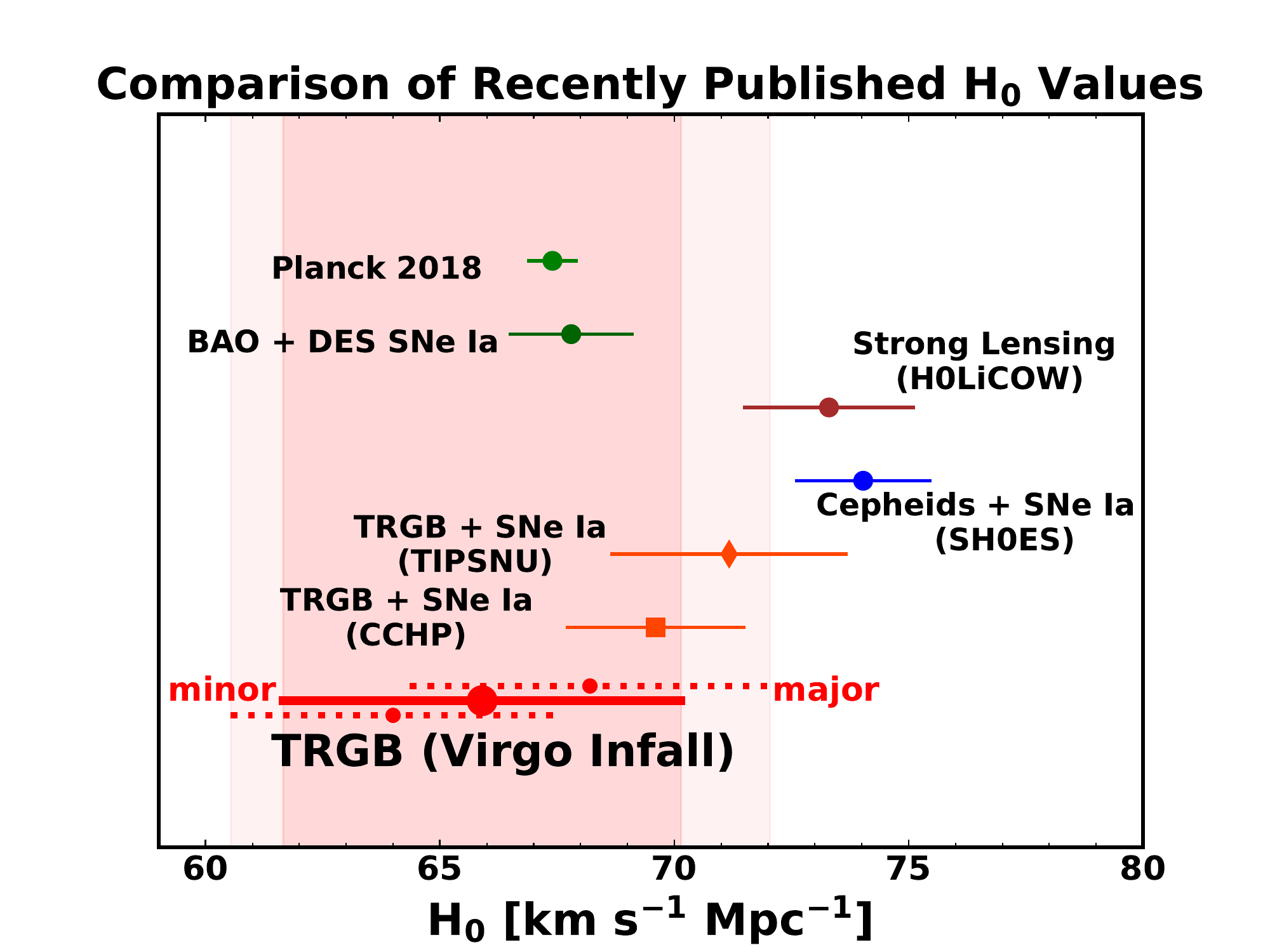} 
\caption{Comparison of $H_0$ obtained in this work (large red circle is the average, and small red circles are the minor and major attractor models) with other recent determinations (CMB: \citet{Planck18}; DES SNe Ia + BAO: \citet{BAO}; Strong Lensing: \citet{LENSING}; Cepheids + SNe Ia: \citet{Riess19}; TRGB + SNe Ia (TIPSNU): \citet{jan17b}, diamond symbol; TRGB + SNe Ia (CCHP): \citet{freed20}, square symbol)  
The horizontal bars indicate 1$\sigma$ error range.
The darker shaded region shows the 1$\sigma$ error range of $H_0$ from the average of minor and major attractor models and the lighter shaded region shows 1$\sigma$ ranges of $H_{0, {\rm minor}}$ and $H_{0, {\rm major}}$. }
\label{fig_H0}
\end{figure*}

We estimate a local Hubble constant $H_0 =$ 65.8 $\pm$ 3.5($stat$) $\pm$ 2.4($sys$)  \kmsMpc~, with 6\% uncertainty using 33 galaxies located between the Virgo cluster and the Local Group.
In this Section, we first compare our $H_0$ result with those of previous studies using Virgo infall.
Then we compare our $H_0$ with those from other recent methods from the literature.

\citet{pei06} fitted 27 galaxies in front of the Virgo cluster with Tully-Fisher distances \citep{tee92} to the velocity-distance relation and obtained $H_0 =$ 65 $\pm$ 9 \kmsMpc, which was revised in \citet{pei08} to $H_0 =$ 71 $\pm$ 9 \kmsMpc.
This value is consistent with the one in this study.
However, it is noted that, with increased accuracy in distance measurements in this study, $H_0$ is determined with a much smaller error, about half of the previous studies.

More recently, \citet{kar18} fitted 28 galaxies with TRGB distances to the velocity-distance model by \citet{pen14, pen17}.
The velocity-distance model of \citet{pen14} has the form for $t_0$ (the age of the universe): 
\begin{equation}\label{eq:pen}
    v = (1.2 + 0.31 \Omega_\Lambda) {{R} \over {t_0}} - 1.1 \sqrt{{{GM} \over {R}}}
\end{equation}
If we set fixed $n =0.5$ when fitting the numerical solution in Eq. \eqref{eq:infall} instead of setting it as a free parameter, our solution is expected to be similar to the above form. 
\citet{kar18} modified this model by dropping the coefficients: $v = H_{0, {\rm Kar}} R - H_{0, {\rm Kar}} \sqrt{ {{R_0^3} / {R}}}$.
This resulted in a large value of $H_{0, {\rm Kar}}$, 97 \kmsMpc~ for the minor attractor model and 104 \kmsMpc~ for the major attractor model.
They did not present the uncertainties of these values. 
Conversion of several variables and using $\Omega_\Lambda =$ 0.685 in Eq. \eqref{eq:pen} yields $v = 1.48 H_0 R - 1.50H_{0} \sqrt{ {{R_0^3} / {R}}}$, so that $H_{0, {\rm Kar}} \sim 1.49 H_0$.
Then the large Hubble constant in \citet{kar18} reduces to $H_0 =$ 65 \kmsMpc~ for the minor attractor model and 70 \kmsMpc~ for the major attractor model, which are coincident with our results.

Moreover, our $H_0$ agrees well with the very local value of $H_0 =$ 67 $\pm$ 5 \kmsMpc~ determined by \citet{pen14} using the Local Group infall.
They fitted velocities and TRGB distances of 34 galaxies in the Local Group to the theoretical infall model using Bayesian techniques in order to obtain cosmological parameters.

Finally, 
Figure \ref{fig_H0} shows a comparison of our $H_0$ measurement with those based on other methods in the literature.
First, our value of $H_0$ is consistent with the values derived from TRGB-calibrated SNe Ia \citep{jan17b,freed20}. This is remarkable, because they are derived from two totally independent methods based on the same TRGB distances.
Second, our value of $H_0$ is consistent with the values of Planck \citep{Planck18} and baryon acoustic oscillations (BAO) \citep{BAO}. Thus, our measurement shows little tension with the Planck value.
Third, our value of $H_0$ is smaller than the value from Cepheid-calibrated SNe Ia \citep{Riess19}, with the difference being at the level of 
1.8$\sigma$.

\section{Summary}\label{sec_summary}

We obtained TRGB distances to NGC 4437 and NGC 4592 using resolved RGB stars present in the {\it HST} images.
Then we compiled TRGB magnitudes of 33 galaxies located between the Local Group and the Virgo cluster, including our own measurements of NGC 4437 and NGC 4592.
We determined the value of $H_0$ by fitting the distances and velocities of these 33 galaxies to the theoretical velocity-distance relation.
Our main results are summarized as follows.

\begin{itemize}

\item We applied three different methods of TRGB detection to LFs of NGC 4437 and NGC 4592: the GLOESS method \citep{hatt17}, the direct edge detection method \citep{jan17}, and the maximum likelihood method \citep{maka06}. All the measurements with different methods coincided well within their quoted uncertainties. The resulting TRGB distances are 9.28 $\pm$ 0.39 Mpc and 9.07 $\pm$ 0.27 Mpc for NGC 4437 and NGC 4592, respectively.

\item The spatial separation between NGC 4437 and NGC 4592 is 0.29$^{+0.30}_{-0.03}$ Mpc, implying that they are indeed a physical pair, consisting of a galaxy group. It is noted that they are located near the zero-velocity surface of the Virgo cluster.

\item Including our own measurements, we compiled TRGB magnitudes of 33 galaxies located between the Local Group and the Virgo cluster. We found that the local Hubble flow perturbed by the Virgo cluster is well described by a velocity-distance relation as a function of cosmological constants ($H_0$, $\Omega_\Lambda$) and $R_0$. By fitting velocities and distances of the galaxies to the relation using MCMC, we obtained $H_0 =$ 65.8 $\pm$ 3.5($stat$) $\pm$ 2.4($sys$) \kmsMpc~ and $R_0 =$ 6.76 $\pm$ 0.35 Mpc.

\item Our $H_0$ is in agreement with those obtained from TRGB-calibrated SNe Ia \citep{jan17b, freed20}. Moreover, our local $H_0$ is consistent with the Planck value \citep{Planck18} measured from CMB. It is slightly smaller than the value from Cepheid-calibrated SNe Ia \citep{Riess19}.

\item Our estimation of the mass of the Virgo cluster within the zero-velocity surface is M(R$_{0}$) = (5.7 $\pm$ 1.5) $\times$ 10$^{14} M_{\odot}$, which is very similar to the virial mass. This confirms the conclusions of previous studies \citep{kar14, kar18, kas18b, kas20} that the outskirts of the Virgo cluster do not contain a significant amount of dark matter.

\item The Hubble flow around the Virgo cluster is cold, as indicated by small values of rms velocity error of the fit and intrinsic velocity dispersion ($\sigma_v =$ 62 $\pm$ 14 \kms).

\end{itemize}

\bigskip
We thank the anonymous referee for useful suggestions. This  work  was supported by the National Research Foundation grant funded by the Korean Government (NRF-2019R1A2C2084019). 
J.K. was supported by the Global Ph.D. Fellowship Program (NRF-2016H1A2A1907015) of the National Research Foundation (NRF).
We thank Brian S. Cho for his help in improving the English in this paper.

\clearpage

\end{document}